\documentclass[structabstract]{aa}  
\usepackage{graphicx}
\usepackage{natbib}
\usepackage{txfonts}
\begin{document}
   \title{The mean infrared emission of Sagittarius\,A*}

   \author{R. Sch\"odel
          \inst{1}
          \and
          M. R. Morris
          \inst{2}
          \and
          K.Muzic
          \inst{3}
          \and
          A. Alberdi
          \inst{1}
          \and 
          L. Meyer
          \inst{2}
          \and
          A. Eckart
          \inst{4}
          \and
          D. Y. Gezari
          \inst{5}
          }

   \institute{Instituto de Astrof\'isica de Andaluc\'ia (CSIC),
     Glorieta de la Astronom\'ia s/n, 18008 Granada, Spain\\
              \email{rainer@iaa.es}
              \and
             UCLA Division of Astronomy and Astrophysics, Physics and
              Astronomy Building, 430 Portola Plaza, Box 951547, Los
              Angeles, CA 90095-1547, USA
             \and
               Department of Astronomy and Astrophysics, University of
               Toronto, 50 St. George Street, M5S 3H4 Toronto ON, Canada
              \and 
              I. Physikalisches Institut, Universit\"at zu K\"oln,
              Z\"ulpicher Str. 77, 50937 K\"oln, Germany
             \and
             NASA/Goddard Space Flight Center, Code 667, Greenbelt, MD
             20771, USA
%             \thanks{The university of heaven temporarily does not
%                     accept e-mails}
             }

   \date{}

% \abstract{}{}{}{}{} 
% 5 {} token are mandatory
 
  \abstract
  {The massive black hole at the center of the Milky Way,
    Sagittarius\,A* (Sgr\,A*) is, in relative terms, the weakest
    accreting black hole accessible to observations. It has inspired the
    theoretical models of radiatively inefficient
    accretion. Unfortunately, our knowledge of the mean SED and
    source structure of Sgr\,A* is very limited owing to numerous observational
    difficulties. At the moment, the mean SED of Sgr\,A* is only known reliably in
    the radio to mm regimes.}
   {The goal of this paper is to  provide constraints on the mean
     emission from Sgr\,A* in the near-to-mid infrared.}
   {Sensitive images of the surroundings of Sgr\,A* at $8.6\,\mu$m,
     $4.8\,\mu$m , and $3.8\,\mu$m were produced by combining large
     quantities of imaging data. Images were produced for several
     observing epochs. Excellent imaging quality was reached in the MIR
     by using speckle imaging combined with holographic image
     reconstruction, a novel technique for this kind of data. }
   {No counterpart of Sgr\,A* is detected at $8.6\,\mu$m. At this
     wavelength, Sgr\,A* is located atop a dust ridge, which
     considerably complicates the search for a potential
     point source. An observed $3\,\sigma$ upper limit of
       $\sim$$10$\,mJy is estimated for the emission of Sgr\,A* at
       $8.6\,\mu$m, a tighter limit at this wavelength than in
       previous work. The de-reddened $3\,\sigma$ upper limit,
       including the uncertainty of the extinction correction, is
       $\sim$$84$\,mJy . Based on the available data, it is argued that,
     with currently available instruments, Sgr\,A* cannot be detected
     in the MIR, not even during flares. At $4.8\,\mu$m and
     $3.8\,\mu$m, on the other hand, Sgr\,A* is detected at all times
     at least when considering timescales of a few up to 13 min. We
     derive well-defined time-averaged, de-reddened flux densities of
     $3.8\pm1.3$\,mJy at $4.8\,\mu$m and $5.0\pm0.6$\,mJy at
     $3.8\,\mu$m. Observations with NIRC2/Keck and NaCo/VLT from the
     literature provide good evidence that Sgr\,A* also has a fairly
     well-defined de-reddened mean flux of $0.5-2.5$\,mJy at
     wavelengths of $2.1-2.2\,\mu$m.  }
   { We present well-constrained anchor points for the SED of Sgr\,A*
     on the high-frequency side of the Terahertz peak. The new data
     are in general agreement with published theoretical SEDs of the
     mean emission from Sgr\,A*, but we expect them to have an
     appreciable impact  on the model parameters in future theoretical work.}

   \keywords{Galaxy: center; Infrared: general}

   \maketitle
%
%________________________________________________________________

\section{Introduction}

Sagittarius\,A* (Sgr\,A*) is the name of the electromagnetic source
related to the massive black hole at the center of the Milky Way. It
is located at a distance of about 8\,kpc and has a mass of roughly 4
million solar masses
\citep[e.g.,][]{Ghez:2008fk,Gillessen:2009qe}. That its
bolometric luminosity amounts to only about $10^{-9}$ times its
Eddington luminosity makes Sgr\,A* -- in relative terms -- the weakest
accreting black hole currently accessible to observation. It is
therefore of great interest for developing and testing theories of
accretion and emission in the weak accretion limit. Sgr\,A* was key to
the development of theories of radiatively inefficient accretion, such as
the ADAF, ADIOS, jet-ADAF, RIAF, CDAF, or similar models \citep[see,
e.g., review by][]{Quataert:2003uq}. An essential feature of all these
models is their success in explaining the extremely low efficiency of
converting accretion power into electromagnetic radiation, which is
for weakly accreting sources like Sgr\,A* many orders of magnitude
lower than in the case of the standard thin accretion disk.  See the
recent reviews by \citet{Melia:2001fk} and \citet{Genzel:2010fk} for
references and details.

Observationally, Sgr\,A* is a difficult target for a combination of
causes. In the regime of gamma and X-rays, the target is relatively
faint, angular resolution is comparatively low, and confusion with
other point or extended sources can be strong.  Strong extinction
makes observations in the optical to ultraviolet/soft X-ray domains
all but impossible. In the near-infrared, Sgr\,A* is faint and
confused with the extremely dense surrounding star cluster. In the
mid-and far-infrared, it has so far eluded detection because of the
high thermal background and confusion with interstellar dust emission. Sensitive
observations with high angular resolution are only just now becoming
possible in the Terahertz regime, but are technically highly
challenging. At centimeter wavelengths, finally, the source is
broadened significantly by interstellar scattering. The intrinsic
source structure can be resolved with very long baseline
interferometry at mm to submm wavelengths
\citep[e.g.,][]{Doeleman:2008fk}, but there are still large technical
difficulties.  All these factors together mean that even almost four
decades after its detection \citep{Balick:1974fk}, we still do not
know the intrinsic source structure of Sgr\,A*, and there are still
gaps of several orders of magnitude in its observed spectral
  energy distribution (SED).

An additional aspect that must be taken into account when examining
Sgr\,A* is that the source is highly variable at wavelengths shorter
than a few millimeters. Particularly in the near-infrared (NIR) and
X-ray domains, Sgr\,A* shows episodes of significantly increased
emission. These so-called {\it flares} last on the order of 100\,min,
with significant substructure ({\it subflares}) on timescales of
$\sim$20\,min. The rise and fall times of flares and subflares can be
as short as a few minutes \citep[see,
e.g.,][]{Baganoff:2001kx,Genzel:2003hc,Ghez:2004fx,Eckart:2008fk,Meyer:2008fk,Porquet:2008kx,Yusef-Zadeh:2008fk,Eckart:2009wm,Yusef-Zadeh:2009fk,Dodds-Eden:2011fk}. Most
of the time, however, Sgr\,A* is very weak at both X-rays and in the
near-infrared. Both infrared observational and theoretical
  efforts have been focused almost exclusively on the flares so far.
  An exception are the works by \citet{Do:2009ij} and
  \citet{Dodds-Eden:2011fk}, which are not biased toward flare emission
  and show that Sgr\,A* is continuously variable at wavelengths of
  $2.1-2.2\,\mu$m (in the so-called $K-$band).  While the analysis of
  the short-timescale variability is of great interest for the
  emission and accretion processes or even potentially for tests of
  general relativity
  \citep[e.g.,][]{Broderick:2006fk,Meyer:2006uq,Paumard:2008fk,Zamaninasab:2011fk},
  it is principally the mean, time-averaged emission that is of
  interest for models of the accretion/outflow processes that give
  rise to the overall SED of Sgr\,A*
  \citep[e.g.,][]{Yuan:2002uq,Moscibrodzka:2009kx}. Since there may be
  up to a few bright flares per day, these timescales range from one
  to several days. In this work, we focus on the mean
emission of Sgr\,A* in the infrared regime.

The goal of this paper is to improve on available measurements of the
mean emission from Sgr\,A*, and in particular, to tighten the constraints
at $8.6\,\mu$m and to measure its mean flux-density at $3.8$ and
$4.8\,\mu$m.

Sgr\,A* is located within the so-called mini-spiral, a prominent
feature of the interstellar medium in the central parsec of the
Galactic center (GC) that is bright at mid-infrared (MIR) wavelengths,
probably due to emission from warm dust \citep[see, e.g., reviews
by][]{Morris:1996vn,Mezger:1996uq,Genzel:2010fk}. High angular
resolution is therefore required to separate Sgr\,A* from the
surrounding complex ISM emission. \citet{Stolovy:1996uq} reported that
Sgr\,A* sits on a ridge of emission, from their deconvolved images.
Later observations showed the ridge much more clearly and that there
is no obvious point-source corresponding to Sgr\,A*, which could be
separated from the ridge \citep{Schodel:2007hh}. All attempts to find
a MIR counterpart of Sgr\,A* have been unsuccessful so far, even
during NIR flares \citep[most recent work
by][]{Eckart:2006sp,Schodel:2007hh,Dodds-Eden:2009mi}. 
  Extinction-corrected $3\,\sigma$ upper limits on Sgr\,A* so far in
  the MIR are reported as $64$\,mJy at $8.59\,\mu$m
  \citep{Schodel:2007hh} and $57$\,mJy at $11.88\,\mu$m
  \citep{Dodds-Eden:2009mi}\footnote{These upper limits do not
    include the uncertainty of the extinction correction. The
    extinction assumed by \citet{Schodel:2007hh} is probably
    under-estimated}.

As concerns the $M'$-band, we are only aware of two publications
reporting on Sgr\,A* at $4.8\,\mu$m. \citet{Clenet:2004ve} reported on
$M'$ observations of a Sgr\,A* counterpart with NaCo/VLT at two
epochs, but it is not clear whether they could distinguish clearly
between Sgr\,A* and a nearby dust-blob
\citep{Ghez:2005fk}. \citet{Hornstein:2007kx} report the detection of
a variable Sgr\,A* counterpart with NIRC2/Keck in $M_s$
\footnote{Within the accuracy required for this work, we can neglect
  the difference between NaCo $M'$, $\lambda_{\rm central}=4.78\,\mu$m
  and $\Delta\lambda=0.59$, and NIRC2 $M_s$, $\lambda_{\rm
    central}=4.67\,\mu$m and $\Delta\lambda=0.24$.}  during one
observational epoch with NIRC2/Keck. In this paper we do not intend to
resolve the variability of Sgr\,A* on short timescales, but infer its
mean luminosity in $M'$ during a large number of epochs and on deep
images that average several observing runs.

High sensitivity and accuracy at $8.6\,\mu$m is reached in this work
by using large quantities of imaging data from many observing
  epochs. Additionally, the novel application of the speckle
  holography technique to MIR imaging data delivers high-Strehl
  images for all epochs  and makes deconvolution unnecessary.

The properties of ISM and stars from imaging of a larger region
between $3.8\,\mu$m and $8.6\,\mu$m, about $20"\times20"$ around
Sgr\,A*, are treated in an upcoming companion paper by
Sch\"odel \& Morris.

The following section describes the data used for this work and their
reduction. We then briefly discuss the ISM emission near Sgr\,A* and
subsequently derive an upper limit on the emission from Sgr\,A* at
$8.6\,\mu$m. Measurements of the mean flux density of Sgr\,A* in the $M$-,
$L-$, and $K$-bands are presented in sections\,\ref{sec:M},
\ref{sec:L}, and \ref{sec:K}\footnote{Based on observations collected
  at the European Organisation for Astronomical Research in the
  Southern Hemisphere, Chile under programmes 071.B-0365, 073.A-0442,
  073.B-0665, 076.B-0715, 077.B-0552, 077.B-0028, 179.B-0261,
  079.B-0084, 079.B-0929, 279.B-5022, 081.B-0648, and 082.B-0952.}. We
proceed to discuss the newly derived data on the mean infrared
emission from Sgr\,A* within the context of models for its overall
SED. In the subsequent section, we argue that it is rather improbable
to detect either the mean emission or flares from Sgr\,A* in the MIR
by standard imaging with current telescopes and instruments. The final
section summarizes our conclusions.

%__________________________________________________________________

\section{Observations and data reduction}

\subsection{Observations with VISIR/VLT at $8.6\,\mu$m \label{sec:obs8.6}}

All mid-infrared (MIR) imaging data used in this work were acquired
with the MIR camera VISIR at the ESO VLT \citep{Lagage:2004fk}. The
PAH\,1 filter was used, with a central wavelength of $8.59\,\mu$m and
a half-band width of $0.42\,\mu$m. Data used in this work are from 5
June 2006, 1/3/5/6/8 April 2007, 22/23 May 2007, and 20/21/23/24 July
2007 (all dates given in UTC)\footnote{Observations in June 2006 and
  May 2007 were done by the first author, the other data were obtained
  from the ESO Science Archive}. The pixel scale was $0.075"$ per
pixel. Conventional MIR imaging was used in the observations from 5
June 2006, with standard data reduction, as described in
\citet{Schodel:2007hh}, who used the same June 2006 data. During the
selected observing runs in 2007, the so-called {\it burst mode} of
VISIR was used \citep{Doucet:2006fk} to record individual
short-exposure frames and the images were reconstructed with a speckle
holography algorithm
\citep[see][]{Liu:1973uq,Bates:1973kx,Petr:1998vn}

In the standard imaging mode of VISIR, several tens of exposures (with
detector integration times around 20\,ms) are averaged by the camera
electronics before they are stored on disk. The burst mode suppresses
the averaging and allows one to store the individual frames and thus
to use VISIR as a speckle camera. Speckle image reconstruction methods
can subsequently be applied to the burst mode data, which can lead to
significant improvement of the image quality
\citep[see][]{Doucet:2006fk} because the deteriorating effects of
atmospheric turbulence can be partially compensated by the
reconstruction process. A pre-requisite for this technique is the
presence of a sufficiently bright ($\sim5$\,Jy) point source reference
in the FOV.  For example, a typical data reduction method would be to
apply a simple shift-and-add (SSA) procedure \citep[see,
e.g.,][]{Christou:1991kx} to the individual frames, possibly combined
with frame selection. The result is an increased Strehl ratio of the
final image compared to a traditional long-exposure image.

\begin{figure}[!htb]
\includegraphics[width=\columnwidth,angle=0]{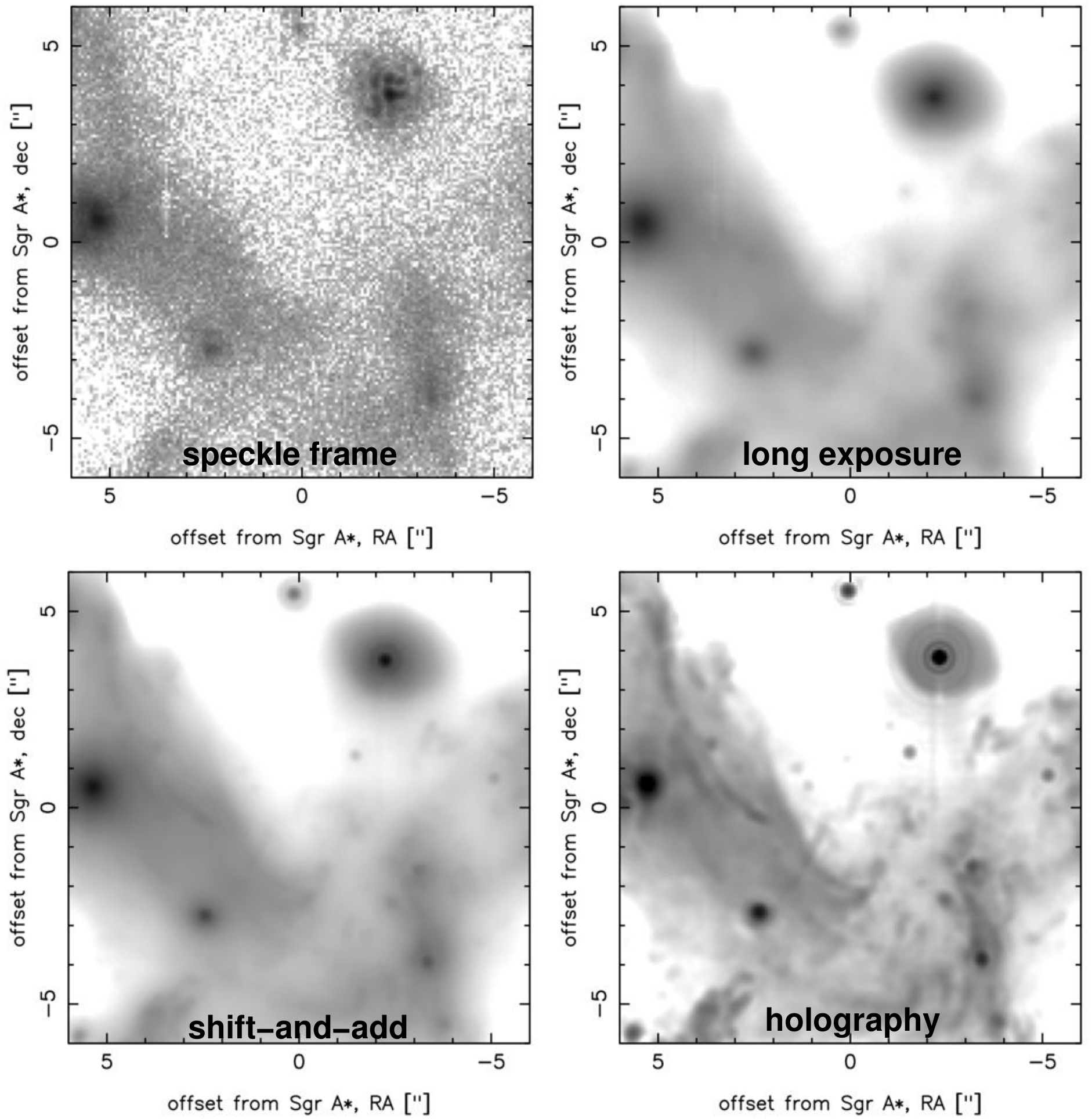}
\caption{\label{Fig:may07} Speckle imaging with VLT/VISIR at
  $8.59\,\mu$m; data from 21/22 and 22/23 May 2007.  Upper left:
  Individual speckle frame. Upper right: Long exposure image,
  i.e. straight average image with only a tip-tilt correction applied,
  similar to standard imaging. Lower left: Simple shift-and-add
  image. Lower right: Image reconstructed with the speckle holography
  algorithm. the vertical line through IRS\,3 visible in the lower
  right panel is an artifact of the detector.}
\end{figure}

The bright star IRS\,3, located about $5"$ NW of Sgr\,A*, was used as
point source reference for the speckle technique. The PSF of a point
source in short exposure (burst mode) images in the MIR is usually
dominated by a single speckle. However, the seeing during the May 2007
observations was so bad (visual seeing $\sim$$2"$ according to the
Paranal DIMM monitor) that IRS\,3 appeared as a cloud of speckles in
the majority of images (see upper left panel in
Fig.\,\ref{Fig:may07}).  A simple shift-and-add (SSA) algorithm only
takes advantage of the brightest speckle in the speckle cloud. The SSA
algorithm can be improved through the selection of the best speckle
frames, i.e. the ones with the highest S/N, which are short exposures
of almost perfect image quality. However, the cost of such a {\it
  lucky imaging} technique is the loss of a large portion of the
frames, typically 50\%-90\%, and thus a low efficiency. As an
alternative, we chose to use the speckle holography technique for
image reconstruction. Speckle holography uses the information in the
entire instantaneous PSF (the speckle cloud) and therefore leads to
images with high Strehl ratios without the need to discard any
significant fraction of frames. A clear and concise description of the
technique is presented in \citet{Petr:1998vn}. We used IRS\,3 as PSF
reference. The fact that IRS\,3 is surrounded by a bow-shock of
diffuse emission \citep[see Fig.\,\ref{Fig:may07} or
][]{Viehmann:2006uq} proved to be no impediment because the surface
brightness of the diffuse flux is a factor $\sim$$100$ fainter than
the flux in the PSF of the point source and disappears within the
noise in the individual speckle frames. The diffuse flux is therefore
effectively suppressed by the noise threshold in our holography
algorithm, that suppresses the noise in the determination of the
instantaneous PSF.

The image reconstructed from the May 2007 burst mode data with the
speckle holography algorithm is shown in the lower right panel of
Fig.\,\ref{Fig:may07}. For comparison, we show a single speckle frame
in the upper left, the (tip-tilt corrected) long exposure image in the
upper right, and the SSA image in the lower left panel. The image from
the holography reconstruction is of extraordinary quality. This
  can be contrasted  with the expectations from seeing-limited
  imaging: Seeing in the visual was about $2"$ during the
  observations. The wavelength-dependence of seeing can be
  approximated with the Roddier formula, as $FWHM\propto \lambda^{-0.2}$, which
  predicts a FWHM $\sim$$1"$ at $8.59\,\mu$m. However, the FWHM of the
  reconstructed image is $0.25"$, the diffraction limit of the
  VLT at the observing wavelength. More than three diffraction rings
  are clearly visible around IRS\,3.

The VISIR burst mode observing template only provided chopping along
the north-south (or east-west) direction; the chop throw was set to
$\sim 15"$.   We subtracted the flux-calibrated (see below) May 2007
image from the June 2006 image to estimate the possible systematic
photometric uncertainty due to the non-ideal chopping. From a region
approximately $5"\times5"$ in size, centered on Sgr\,A*, we estimate
an average uncertainty of the flux in each pixel of $\leq$$10\%$. 

A deep image was created by combining all available burst mode frames
from 2007 with the speckle holography algorithm for image
reconstruction. A total of about $900000$ frames with individual frame
integration times of $0.016$\,s and $0.020$\,s were used,
corresponding to a total integration time on target of roughly
$4.5$\,h.  An issue with some of the April and July 2007 data,
retrieved from the ESO archive, was the small chopping throw of just
$10"$ in direction north-south. We checked whether this caused a
significant flux bias in the area near Sgr\,A*, due to chopping into
point-like or extended sources to the north or south. For this purpose
we examined the difference between the images from just the May 2007
data, which should be free from significant errors due to chopping,
and from all 2007 data. Within $\pm1"$ from Sgr\,A* we found a
negligible flux offset and a standard deviation between the pixel
fluxes of $\leq10\%$. Therefore we can be confident
that the flux calibration of all images reconstructed from the 2007
burst mode data is accurate to within at least $10\%$.

The probably best observations (no chopping into obvious sources,
large chopping throw, excellent seeing) are from 8 April 2007
($\sim$$30,000$\,frames) and are presented separately here because of
their high quality. 

The astrometric reference frame and absolute photometry were
established using the imaging data from 4/5 June 2006. The reasons for
this choice are the optimized chopping angle and throw of these data
as well as the use of multiple dither positions. Additionally, both
chopping and nodding were applied during these observations (only
chopping in case of the burst mode observations). All these factors
contribute to the suppression of systematic effects over a relatively
large FOV.  The \emph{StarFinder} program package
\citep{Diolaiti:2000fk} was used for measuring positions and fluxes of
point sources as well as for obtaining a smooth estimate of the
diffuse emission. Astrometric and photometric mean values and
uncertainties were estimated by using two different PSFs, one
determined from the star IRS\,10EE, the only reasonably bright one in
the MIR image that is not associated with evident diffuse emission,
and one from the standard star HD\,145897 (see below).  The formal
uncertainties calculated by {\it StarFinder} for each source were
added in quadrature to the uncertainty estimated from the use of the
two different PSFs.

The zero point for the $8.59\,\mu$m PAH\,1 filter was measured via
observations of standard stars before, during, and at the end of the
GC observations. The observations of the three standard stars
HD\,198048, HD\,178345, and HD\,145897 followed the ESO VISIR
calibration plan. Zero points were determined using the
\emph{visir\_img\_phot} algorithm of the ESO Common Pipeline Library
(CPL). The maximum deviation between the mean and the individual
measurements was $3.2\%$, indicating very stable atmospheric
transmission during the entire observing time. The deviation from the
zero point determined by \citet{Schodel:2007hh}, who used the same
data but only one of the observed standard stars and a different
algorithm for estimating the zero point, is also just $3\%$.  We adopt
this value of $3\%$ as conservative $1\,\sigma$ uncertainty of the
absolute photometric calibration of the PAH\,1 image.

In a next step, the astrometric reference frame was established, i.e.,
pixel positions of stars were converted into offsets in arc-seconds
from Sagittarius\,A*. To take possible optical distortion across the
image into account, a polynomial with terms up to second order was
used. The coefficients of the polynomial were computed via
least-squares minimization between the pixel positions of 9 stars
detected in the MIR image (IRS\,15NE, IRS\,7, IRS\,3, IRS\,29N,
IRS\,6E, IRS\,12N, IRS\,9, IRS\,10EE, IRS\,17) and known near-infrared
positions and proper motions of the stars relative to Sgr\,A*, as
given in \citet{Schodel:2009zr}. The uncertainty of this
transformation was determined with a Monte Carlo simulation.  The
astrometric positions of the reference stars were varied randomly in
100 tries within their 1\,$\sigma$ combined NIR and MIR positional
uncertainties. The standard deviation of the resulting position for
each star in the MIR frame was then adopted as the 1\,$\sigma$
uncertainty of its astrometric position and added quadratically to the
uncertainty of its position in the field as measured by
\emph{StarFinder}.

A comparison of the positions of the maser stars (IRS\,7, IRS\,10EE,
IRS\,15NE, IRS\,9, IRS\,17) with the respective ones given in
\citet{Schodel:2009zr} showed a coincidence within the respective
$1\,\sigma$ uncertainties.  The position of Sgr\,A* in the MIR image
{\it relative} to prominent sources, like IRS\,16NW or IRS\,29, could
be determined with an uncertainty of $0.04$\,pixel or about
$3$\,mas. The \emph{absolute} astrometric uncertainty of the
stellar positions in the MIR image will be higher due to the absolute
uncertainty of the NIR reference system used here. As illustrated in
Fig.\,5 of \citet{Schodel:2009zr}, the absolute astrometric
uncertainty is thus on the order of $\sim$$3$\,milliarcseconds (mas) near
Sgr\,A*, but as large as 50\,mas at $\sim$15$"$ distance from Sgr\,A*.

The flux density scale in all other images (from the 2007 data) was
calibrated by using the fluxes of the sources IRS\,2L, IRS\,7, and
IRS\,13E as measured in the June 2006 image.  The uncertainty of this
procedure was estimated from the deviations of the fluxes of the three
reference stars after calibration from their initially assumed
fluxes. The uncertainty was found to be about 10\%. This uncertainty
was added quadratically to the uncertainties of the PSF fitting.  The
FOV of the 2007 imaging data is smaller than the one of the 2006
data. Therefore, absolute astrometry was established via the
astrometric positions of the sources IRS\,7, IRS\,3, IRS\,29N,
IRS\,6E, IRS\,21, and IRS\,1W \citep{Schodel:2009zr}. The position of
Sgr\,A* in all images was determined with a $1\,\sigma$ uncertainty of
$0.04$\,pixels, or 3\,mas.  We detected no significant systematic
differences between the photometric and astrometric calibration of the MIR
images in this work and in our previous work \citep{Schodel:2007hh}.

\subsection{Observations with NaCo/VLT at $4.8\,\mu$m \label{sec:MpObs}}

The central $\sim$$15"\times15"$ around Sgr\,A* were observed in $M'$
with the near-infrared camera and adaptive optics system NAOS/CONICA
\citep[short: NaCo, see][]{Lenzen:2003fk,Rousset:2003uq} at the ESO
VLT at several epochs in 2003, 2004, and 2006
(Tab.\,\ref{Tab:MpSgrA}). The data were obtained from the ESO science
archive.  Overall observing efficiency was low (due to chopping,
dithering, read-out overheads, and possible other reasons unknown to
the authors), so that the total on-source integration time of the
$M'$-band images was between half a minute and two minutes per
observing epoch (Tab.\,\ref{Tab:MpSgrA}). Astrometry was established
via 13 sources in the FOV, using stellar positions and proper motions
from \citet{Schodel:2009zr}. The resulting $1\,\sigma$ uncertainty of
the pixel position of Sgr\,A* in the $M'$-images is
$0.05-0.08$\,pixels ($1.2-2.2$\,mas) in both axes.

Photometry and astrometry of point sources in the $M'$-image were
performed with {\it StarFinder}.  Absolute astrometry was established
with the same sources as for the MIR observations.  The absolute
astrometric uncertainty is estimated to be better than $0.05"$
\citep{Schodel:2009zr}. Since Sgr\,A* is clearly identifiable in $M'$,
very high astrometric precision is not a real issue at this wavelength
\citep[see section\,\ref{sec:M} and ][]{Hornstein:2007kx}.

No measurement of the NACO $M'$ zero point could be found on the ESO
archive. Therefore, the stars IRS\,16C and IRS\,16NW were used to
calibrate the zero point of the $M'$ image. Their magnitudes are
$L'=8.20\pm0.15$ and $PAH\,I=8.05\pm0.1$ for IRS\,16C, and
$L'=8.43\pm0.15$ and $PAH\,I=8.67\pm0.1$ IRS\,16NW. The $L'$
magnitudes are from \citet{Schodel:2010fk} and the $PAH\,1$ magnitudes
from photometry on the 2006 $8.6\,\mu$m data (see previous section).
We assume that extinction stays approximately constant between the
$L'$-band and $8.6\,\mu$m \citep[see][]{Lutz:1999yf,Nishiyama:2009oj}.
It appears that both stars do not show any significant MIR excess and
can be approximated as black bodies. We therefore assume the same
magnitude in $M'$ as in $L'$ for both stars and use them to calibrate
the $M'$ image. The systematic 1\,$\sigma$ uncertainty of the $M'$
calibration is $\sim0.15$\,mag and results from the combination of an
estimated $0.1$\,mag uncertainty in extinction and $0.1$\,mag
uncertainty from the $L'$ calibration (from the average of two
sources).  \citet{Hornstein:2007kx} used an independent calibration
procedure and obtained for IRS\,16C and IRS\,16NW $M_S=8.08$ and
$M_S=8.41$, respectively. We use as flux density zero point 160\,Jy
\citep{Cohen:1992fk}, they used 163\,Jy
\citep{Tokunaga:2005fk}.

\begin{table}[htb]
\caption{Flux density of $M'$ counterpart of Sgr\,A*. \label{Tab:MpSgrA}} 
%\centering
\begin{tabular}{llll}
\hline
\hline
UTC Date & UTC Time & $t_{\rm on\,source}$\tablefootmark{a} & $f_{\rm SgrA*}$\tablefootmark{b}\\
\hline
2003 Jun 04 & 04:21 - 07:36 & 130  & $1.6\pm0.2$ \\
2003 Jun 09 & 04:29 - 07:20 & 115 &$1.2\pm0.2$ \\
2004 Jun 14 & 03:30 -05:56  &  58 & $1.8\pm0.3$ \\
2004 Aug 11 & 00:54 - 01:15 & 27 &$1.8\pm0.4$ \\
2004 Sep 20 & 00:06 - 01:47  & 55 &$1.6\pm0.2$ \\ 
2004 Sep 20/21 &  23:29 - 00:30  & 95& $1.2\pm0.2$ \\
2006 Mar 28 & 07:11 - 08:49 & 47 & $1.2\pm0.3$ \\
2006 Mar 29 & 07:06 - 10:24 & 122 &$1.8\pm0.2$ \\ 
all 2003\tablefoottext{c} & & 245 & $1.4\pm0.2$ \\
all 2004\tablefoottext{c} & & 235 & $1.6\pm0.2$ \\
all 2006\tablefoottext{c} & & 169 & $1.6\pm0.2$ \\
\hline
\end{tabular}
\tablefoot{
\tablefoottext{a}{Total integration time on source in seconds.}
\tablefoottext{b}{Flux density and $1\,\sigma$ uncertainties in mJy, not corrected for extinction.}
\tablefoottext{c}{Measurement on average image from all 2003, 2004, or
2006 imaging data, respectively.}
}
\end{table}

\begin{figure}[!tb]
\includegraphics[width=\columnwidth,angle=0]{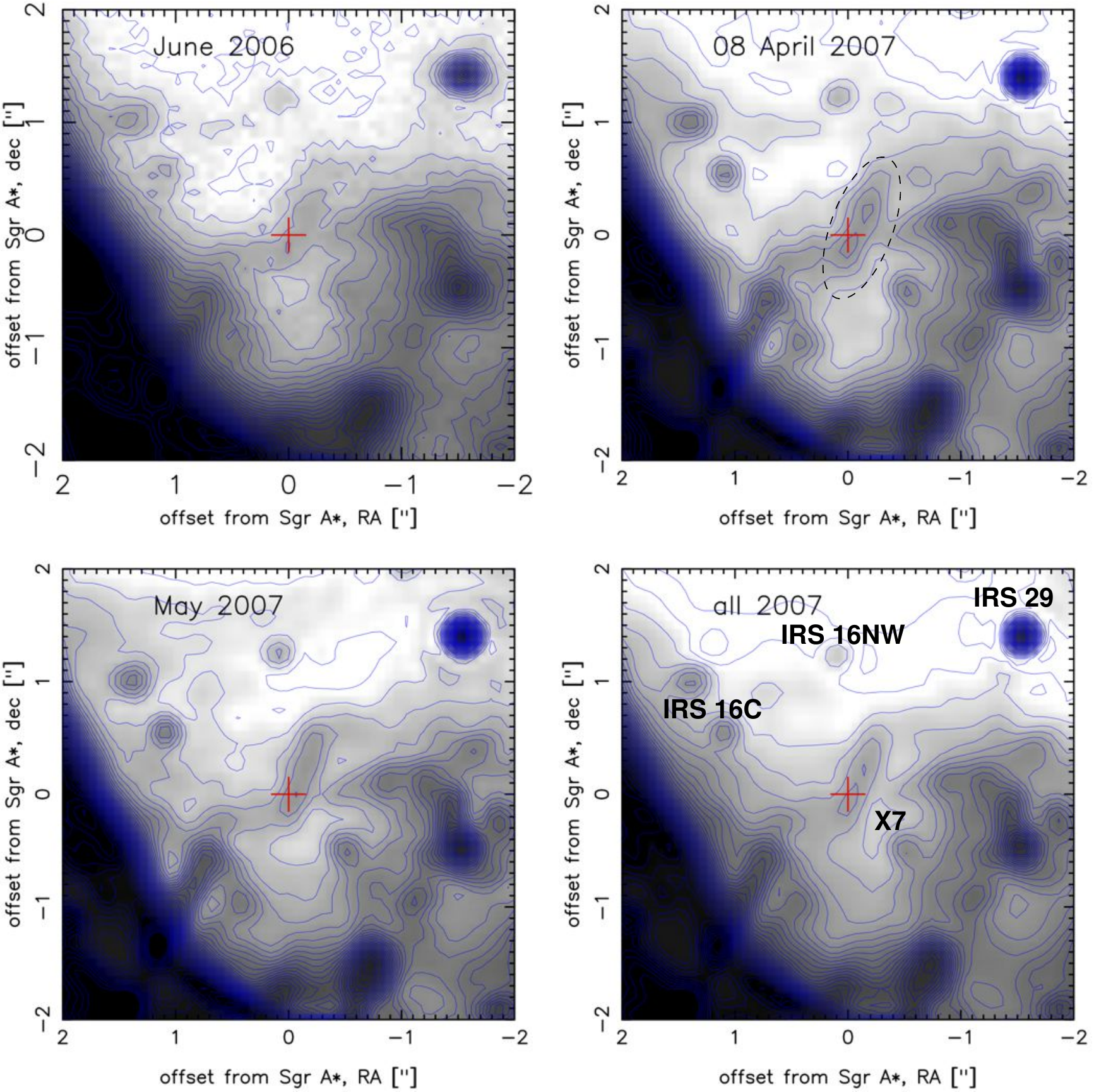}
\caption{\label{Fig:overview} The surroundings of Sgr\,A* at
  $8.6\,\mu$m. Upper left: Image from long-exposure observations on 5
  June 2006. Upper right: Image reconstructed with the speckle
  holography algorithm from burst mode observations on 8 April
  2007. Lower left: Image reconstructed with the speckle holography
  algorithm from burst mode observations on 22 and 23 May 2007. Lower
  right: Image reconstructed with the speckle holography algorithm
  from all burst mode observations in 2007.  North is up and East is
  to the left. Contour lines are plotted in steps of $0.5$\,mJy from
  $0.5$ to $20$\,mJy per pixel (one pixel corresponds to
  $0.075"\times0.075"$). Sgr\,A* is marked with a red cross of $0.3"$
  line segments. The stellar sources IRS16\,NW, IRS16\,C, and IRS29, as well
  as the cometary shaped source X7 \citep{Muzic:2010fk} are labeled
  in the lower right panel. The {\it Sgr\,A*-Ridge} is marked by a
  dashed ellipse in the upper right panel.}
\end{figure}

\subsection{Observations with NaCo/VLT at $3.8\,\mu$m}

The central parsec around Sgr\,A* has frequently been observed with
NaCo/VLT since 2002. In order to constrain the emission of Sgr\,A* in
the $L'$-band ($3.8\,\mu$m), we downloaded L'-imaging data from the
years 2006-2009. The principal reason to choose this time window is
that S2/S0-2, the brightest star near Sgr\,A*, was relatively far from
the black hole during this period. The small proper motion of S2 near
the apocenter of its orbit also ensures that the star will hardly
appear blurred on images created from averaging the data from
different epochs. No selection criterion was applied to the data other
than imaging quality (atmospheric conditions, AO
performance). Particularly important is that the flux state of Sgr\,A*
was not used as a criterion to include or exclude a particular data
set.

Data reduction was standard (sky subtraction, bad pixel interpolation,
flat-fielding). Epochs of low data quality, e.g., because of strongly
variable transmission, extreme turbulence and consequentially
insufficient AO correction, or strong readout patterns on the
detector, were discarded.  In the data from the remaining epochs,
frames with insufficient AO performance (sometimes the AO loop opens),
insufficient removal of sky-background (due to rapidly variable
background in the L'-band) or otherwise corrupted frames were
de-selected before creating final mosaics. The percentage of
  discarded frames ranges between 0\%-70\%, depending on the observing
  epoch. For AO observations of the GC with NaCo/VLT the
  L'-band is frequently used as fall-back option when atmospheric
  conditions are bad, particularly when seeing is bad and fast. It is
  for this reason that the number of discarded frames is very high for
several epochs. The
selected observing epochs, along with the total exposure time for each
one (excluding de-selected frames) are listed in
Tab.\,\ref{Tab:LpObs}.

Photometry and astrometry of point sources in the $L'$-image were
performed with {\it StarFinder}.  Absolute astrometry was established
with the same sources as for the MIR observations.  The absolute
astrometric uncertainty is estimated to be better than $0.05"$
\citep{Schodel:2009zr}. Since Sgr\,A* is clearly identifiable in $L'$,
very high precision in astrometry is not a real issue at this
wavelength. Photometric calibration is based on the magnitudes of
IRS\,16C and IRS\,16NW provided by \citet{Schodel:2010fk}.

\begin{table*}[htb]
\caption{Observations of the GC in the $L'$-band with NaCo, used in
  this work. The detector integration time (DIT) was $0.2$\,s in all
  cases, except 28 Mar 2006, when it was $0.175$\,s.\label{Tab:LpObs}} 
%\centering
\begin{tabular}{lllllll}
\hline
\hline
UTC Date & UTC Time  & good data\tablefootmark{a} & t$_{\rm
  on\,source}$\tablefootmark{b} & f$_{\rm mean,  SgrA*}$\tablefootmark{c} & f$_{\rm median, SgrA*}$\tablefootmark{d} & f$_{\rm S2}$\tablefootmark{e}\\
\hline
2006 Mar 28 & 05:60 - 10:29 & 0.74  &850.5 & 0.64 &  0.82  & 1.47  \\
2006 May 29 & 04:39 - 10:51 & 0.45 &2400 & 1.21  &   0.97  & 1.86 \\
2006 May 30 & 06:05 - 10:24 & 0.52&2400 & 1.72  &   1.62    & 1.96 \\
2006 May 31 & 08:29 - 10:37 & 0.76 &2490 & 1.65  &    1.65   & 2.07 \\
2006 Jun 02 & 04:47 - 07:01 &  0.53&1890    & 2.23  &   2.10  & 1.88\\
2006 Jun 04 & 05:03 - 10:22 &  0.40 &3360  & 1.86   &   1.85   &  2.10\\
2006 Jun 06 & 04:49 - 10:33 &  0.39 &3360  & 1.11  &   1.04    & 2.02\\
2007 Apr 01 & 05:18 - 06:43 &  0.95 &3420 &  0.92  &    0.90   & 1.85\\
2007 Apr 03 & 04:57 - 06:27 &  1.00&3600 & 1.17  &   1.22   & 1.86 \\
2007 Apr 04 & 04:50 - 06:10 &  0.93 &3000 & 3.14 &     3.08   & 1.66 \\
2007 Apr 05 & 08:01 - 10:35 &  0.88 &4890 & 1.51  &     1.39  &  2.08\\
2007 Apr 06 & 05:19 - 06:55 &  0.89&3210 & 0.72  &    0.74  &  1.68\\
2007 May 23 & 04:39 - 10:36 &  0.26 &3060 & 1.82 &    1.57  &  2.16\\
2007 Jul 20 & 01:35 - 02:11 &     0.94 &1350  & 1.25  &    1.11  & 2.06 \\
2007 Jul 22 & 23:08 - 04:51 &   0.88 &7590 & 1.30  &    1.16  & 1.97 \\
2008 May 26 & 05:43 - 10:38 & 0.60 &3720 & 3.93 &   3.97  & 1.88\\
2008 May 30 & 06:41 - 10:41 & 0.54 &1470  & 1.27 &  1.18   & 1.99\\
2009 Apr 01 & 06:11 - 09:01 &  0.95 &5700 & 1.69 &   1.54  & 1.63\\
2009 Apr 05 & 07:02 - 10:35 &  0.62 &3060  & 0.91 &    0.90 & 1.76\\
\hline
\end{tabular}
\tablefoot{
\tablefoottext{a}{Fraction of data used; rest rejected because of low quality.}
\tablefoottext{b}{Total integration time on source in seconds,
  excluding rejected data.}
\tablefoottext{c}{Measured flux density of Sgr\,A* on mean images, in mJy, not corrected for extinction.}
\tablefoottext{d}{Measured flux density of Sgr\,A* on median images, in mJy, not corrected for extinction.}
\tablefoottext{e}{Measured flux density of S2 on mean images, in mJy, not corrected for extinction.}
}
\end{table*}

\section{The mean emission from Sgr\,A* in the infrared}

\subsection{The immediate environment of Sgr\,A* at $8.6\,\mu$m}

Figure\,\ref{Fig:overview} shows the surroundings within $\pm2"$ of
Sgr\,A* on images reconstructed from data obtained on 5 June 2006,
22/23 May 2007, 8 April 2007, and all 2007. The structures seen on all
images are very similar. The images also resemble closely the ones
shown in Figs.\,1 and 2 of \citet{Schodel:2007hh}, who used the same
June 2006 observations as in this paper and applied Lucy-Richardson
deconvolution. We attribute any obvious differences to the fact that
the images have different signal-to-noise ratios, with the highest S/N
reached in the image reconstructed from the entire 2007 burst mode
data set. Also, the June 2006 image from direct imaging has a lower
Strehl ratio than the images from 2007 that are based on burst mode
observations and reconstructed via the speckle holography
algorithm. The excellent quality of the new images makes
the application of deconvolution unnecessary, thus avoiding any
related potential complications.

As can be seen, Sgr\,A* is not isolated, but located near the middle
of a ridge-like structure, which we term the {\it SgrA*-Ridge}. It is
marked by a dashed oval in the upper right panel of
Fig.\,\ref{Fig:overview}. There is no recognizable point-source
coincident with its position, but there may be a point source that is
confused with diffuse emission.  In order to estimate an upper limit
to the mean flux density of Sgr\,A*, we need a reasonable estimate of the
probable diffuse background flux density at its location.

\subsection{An upper limit on the mean emission from Sgr\,A* at $8.6\,\mu$m}

\begin{figure}[!htb]
\includegraphics[width=\columnwidth,angle=0]{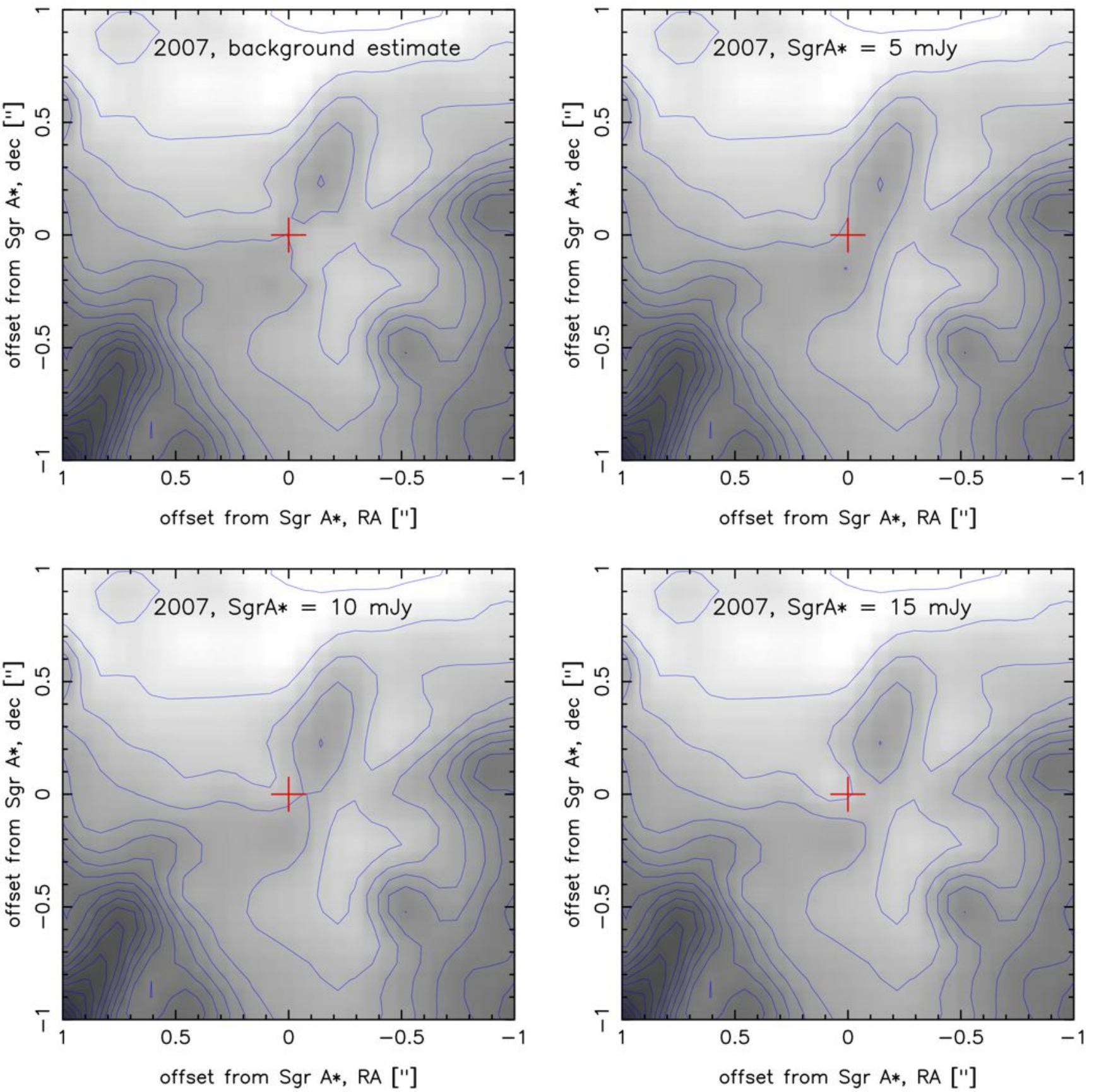}
\caption{\label{Fig:sgrasub} Zoom onto Sgr\,A* at $8.6\,\mu$m in
  the image reconstructed with the speckle holography algorithm from
  all burst mode observations in 2007.  North is up and East is to the
  left. Contour lines are identical to the ones in
  Fig.\,\ref{Fig:overview}. Sgr\,A* is marked with a cross of
  $0.075''$ line segments. Upper left
  panel: Estimate of the diffuse background emission at the position
  of Sgr\,A*. A point source of $5$, $10$, and $15$\,mJy was
  subtracted from the original reconstructed image at the position of
  Sgr\,A* with the results shown in the upper right, lower left,
  and lower right panels, respectively.}
\end{figure}

The upper right, and lower left and right panels of
Fig.\,\ref{Fig:sgrasub} show the image reconstructed from all 2007
data with a point source (convolved with the corresponding PSF) of 5,
10, and 15\,mJy subtracted from the position of Sgr\,A*. At flux densities
$\gtrsim$10\,mJy a depression becomes apparent at the position of
Sgr\,A*. Therefore, we judge that the flux density of Sgr\,A* must be
$\lesssim10\,$mJy.  Another way of estimating the emission from a
putative Sgr\,A* counterpart is by estimating the background emission
at its location and subtracting it from the images.

The background emission at the position of Sgr\,A* can be
  estimated by masking the pixels in a circular area of a few pixels
  radius, centered on Sgr\,A*, and subsequently interpolating the flux
  in the masked pixels with the {\it StarFinder}
  \citep{Diolaiti:2000fk} routine $REPLACE\_PIX$. This routine
  replaces bad pixels with a median of the pixel values in a suitably
  large box around the corresponding location. A minimum of three
  ``good'' pixels is used to compute the median value at a given
  location. For details, see the publicly available source code of
  {\it StarFinder}.  For the 2007 images, a masking radius of 3 (2, 4)
  pixels contains about $78\%$ ($62\%, 80\%$) of the flux on from a
  putative point source at the position of Sgr A*. For a radius
  3\,pixels mask any pixel outside this radius would be contaminated by
  $<5\%$ of the peak pixel flux of such a point source. Therefore,
  this procedure should effectively remove any influence from an
  undiscovered point source at the position of Sgr\,A*. An example of
  the interpolated background is shown in the upper left panel of
  Fig.\,\ref{Fig:sgrasub}, using the image reconstructed from all 2007
  data.  

  The estimated background can subsequently be subtracted from an
  image and the remaining flux at the position of Sgr\,A* can be
  measured and corrected for the used aperture (a stronger correction
  was used for the 2006 image with its lower Strehl ratio).  Due to the emission
  from the {\it Sgr\,A*-Ridge}, the estimated background flux will
  depend on the masking radius. As can be seen in the upper left panel
  of Fig\,\ref{Fig:sgrasub}, a radius 3\,pixels mask will effectively
  remove the ridge emission at the position of Sgr\,A*. A 2 pixels
  masking radius will include more of the ridge emission into the
  background flux at the position of Sgr\,A*, while a 4 pixels masking
  radius will probably under-estimate the background emission at
  the position of Sgr\,A*. We use these three different masking
  radii to measure the remnant flux and estimate its uncertainty.

\begin{table}[htb]
  \caption{Remnant $8.6\,\mu$m flux densities at the position of Sgr\,A* after
    background subtraction. \label{Tab:mirfluxes}} 
%\centering
\begin{tabular}{lll}
\hline
\hline
Date & $f$ & $df$  \\
(UT)  & (mJy) & (mJy) \\
\hline
2006 May 5 & 5.9 & 1.2 \\
2007 Apr 1 & 2.7 & 1.3 \\
2007 Apr 3 & 4.5 & 1.7 \\
2007 Apr 5 & 2.4 & 1.1 \\
2007 Apr 6 & 3.6 & 2.1 \\
2007 Apr 8 & 7.9 & 4.2 \\
2007 May 22 & 7.6 & 3.3 \\
2007 May 23 & 8.3 & 2.5 \\
2007 Jul 20 & 10.1 & 3.9 \\
2007 Jul 21 & 10.1 & 4.2 \\
2007 Jul 23 & 8.8 & 3.6 \\
2007 Jul 24 & 8.3 & 3.4 \\
\hline
\end{tabular}
\end{table}

This procedure was applied to the images from all epochs. The measured
remnant fluxes at the position of Sgr\,A* as well as the corresponding
uncertainties are listed in Tab.\,\ref{Tab:mirfluxes}. The mean of all
12 independent measurements is $6.7\pm0.7$\,mJy, while the weighted
mean is $4.5\pm0.5$\,mJy. However, the measurements from 1, 3, 5, and
6 April 2007 appear to be systematically low. When inspecting those
images, very similar patterns of negative emission can be seen in
large parts of the images, also close to Sgr\,A*. A further
examination of the corresponding imaging data reveals that the
chopping angle was chosen along a N-S direction with an amplitude of
just $8"$. The effect is that bright sources, like IRS\,7, IRS\,4,
IRS\,21, and much extended emission from the northern and southern
parts of the mini-spiral are subtracted from the source images. It is
possible that the Sgr\,A* region and/or the region of the calibrator
sources are affected by these systematic negativities. We therefore
discard these measurements, which is a conservative step because it
will correct the estimated mean flux upwards.  After exclusion of the
problematic April 2007 data sets, we obtain a mean from 8 independent
measurements of $8.4\pm0.5$\,mJy, where the weighted mean is
$7.1\pm0.9$\,mJy.

\citet{Schodel:2007hh} reported $8\pm5$\,mJy as the flux density
  of any putative point source at the position of Sgr\,A*. At first
  glance, there seems to be hardly any difference in the results between
  this work and the earlier one. However mid-infrared observations of
  the Galactic center are non-trivial because of the involved
  systematic uncertainties (see problem of systematic errors related
  to non-optimal chopping described above). Furthermore, there is no
  straightforward way to measure the upper limit of the emission from
  a putative point-like source at the position of Sgr\,A*. The method
  applied here is different and independent from the one applied in
  \citet{Schodel:2007hh}, who subtracted the flux-scaled diffuse
  emission of a point-source subtracted $L'$-band image from the MIR
  image.  The quality and quantity of analyzed data is
  much higher here. The zero point is established with greater
  reliability via the use of three instead of just one standard
  stars. Finally, the use of many different epochs with different
  observing setups (particularly the chopping angle) will minimize the
  influence of systematic errors. The fact that the result here is
  similar to that in our previous work gives us great confidence in the
  estimated upper limit, while the uncertainty has been reduced
  significantly. 

  The measured remnant flux can be regarded as an estimate of the
  upper limit to any potential point-source at the position of
  Sgr\,A*. The deep images produced in this work show no
  indication of any point source at $8.6\,\mu$m at the position of
  Sgr\,A*. It is possible that the dust emission at the position of
  Sgr\,A* reaches levels similar to those measured in the parts of the
  {\it Sgr\,A*-Ridge} immediately NW and SE of Sgr\,A*. In this case,
  the background estimates obtained with a 2 pixel masking radius
  would be most appropriate, resulting in an estimated remnant flux of
  just 3\,mJy.  Since two of the three masking apertures used here are
  larger, we are confident that our measurements are
  conservative. We may even overestimate the remnant flux at
  the position of Sgr\,A* by factors $2-3$.

  Hence, from the weighted mean, we derive a $3\,\sigma$ upper limit
  to the long-term average of the flux density of Sgr\,A* of
  $9.8\,$mJy ($9.9$\,mJy for the unweighted mean). 

  Unfortunately, $\lambda=8.6\,\mu$m is located right on the blue edge
  of the $10\,\mu$m silicate absorption feature, see, e.g.,
  \citet{Moneti:2001fk,Lutz:1996oz}, giving rise to relatively high
  extinction. Here, we use $A_{8.6}\approx2.0\pm0.3$, from the recent
  measurements by \citet{Fritz:2011uq}.  The extinction corrected
  upper limit on the $8.6\,\mu$m emission from Sgr\,A* is thus
  $45\pm13$\,mJy, using the weighted mean.  We note that the
  extinction assumed here is higher than what was used by
  \citet{Schodel:2007hh}. Their upper limit, corrected for the
  extinction used here, is $50\pm33$\,mJy. In the following, we will
  use a de-reddened $3\,\sigma$ upper limit of $84$\,mJy, which
  includes the uncertainty of the extinciton correction.

\subsection{The mean emission from Sgr\,A* at $4.8\,\mu$m \label{sec:M}}

\begin{figure}[!tb]
\includegraphics[width=\columnwidth,angle=0]{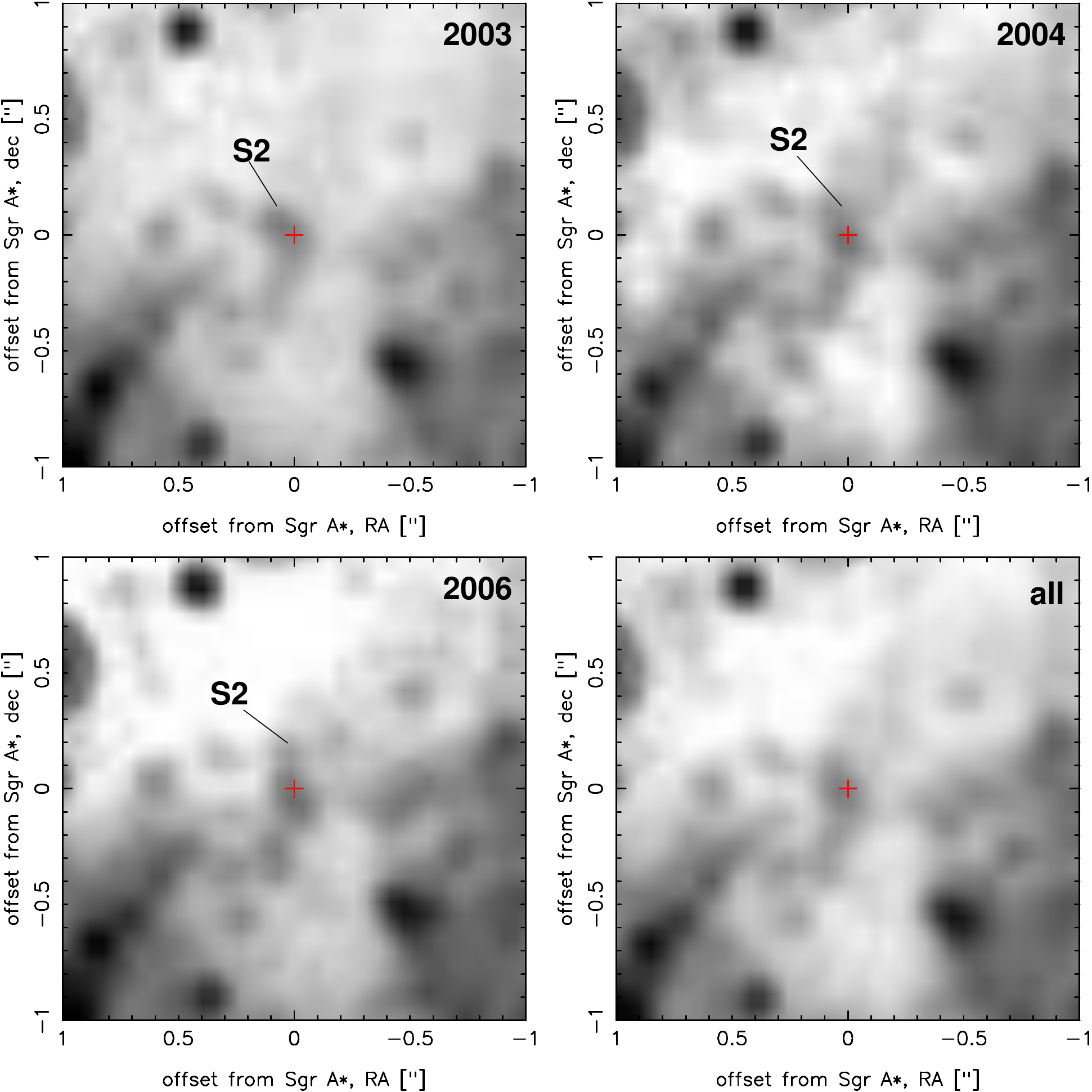}
\caption{\label{Fig:Mp} The surroundings of Sgr\,A* seen with NaCo in
  the $M'$-band. Upper left: Average image from 2003 data. Upper
  right: Average image from 2004 data. Lower left: Average image from
  2006 data. Lower right: Average image from 2003-2006 data.  Sgr\,A*
  is marked with a red cross. The fast-moving star S2 is marked in the
  2003, 2004, and 2006 images. The gray scale is logarithmic.}
\end{figure}

The average images from the 2003, 2004, and 2006 NaCo $M'$-band
imaging data are shown in Fig.\,\ref{Fig:Mp} together with an average
image based on the data from all epochs.  As can be seen, a source can
clearly be detected at the location of Sgr\,A* in the $M'$-band on all
images.  The source near Sgr\,A* appears to be elongated toward the SE
at all epochs. This is likely due to emission from a compact dust blob
located about $0.094"$ to the SE of Sgr\,A*. This extended source was
reported by various authors and has an approximate $L'$-magnitude of
$12.8-13.7$
\citep[e.g.,][]{Clenet:2004ve,Ghez:2005fk,Eckart:2006sp,Hornstein:2007kx}. According
to \citet{Ghez:2005fk}, the dust blob can be modeled as a
two-dimensional Gaussian with a major axis of 120\,mas. This is less
than the diffraction-limited resolution of NaCo at $4.8\,\mu$m
($\sim$125\,mas).

\begin{figure}[!tb]
\includegraphics[width=\columnwidth,angle=0]{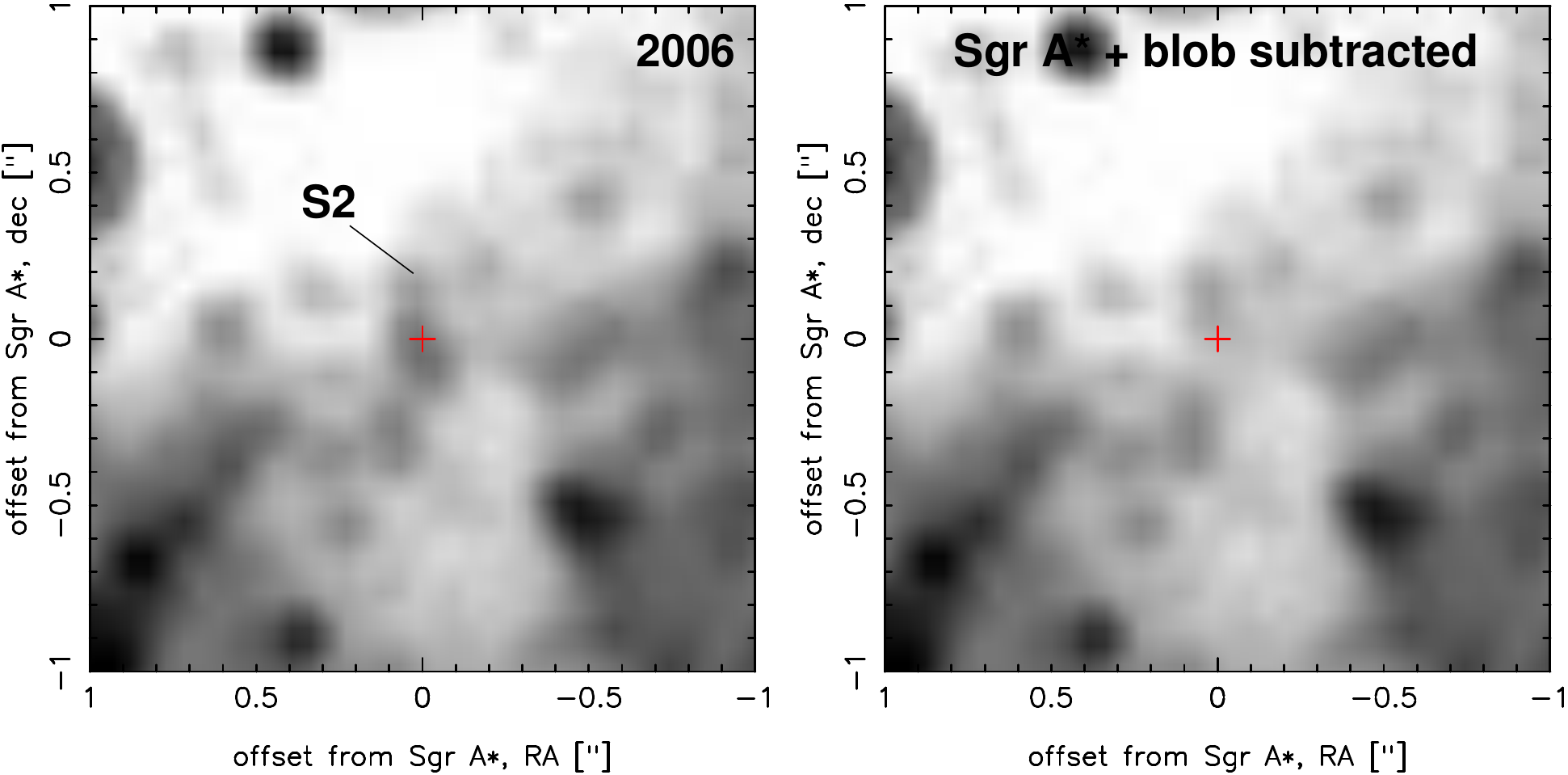}
\caption{\label{Fig:MpSgrA} Left: Average $M'$-image from 2006
  data. Right: The same image, after subtraction of two point sources,
  one at the position of Sgr A* and one $\sim0.094"$ to the SE of
  Sgr\,A*. The position of Sgr\,A* is indicated by a red cross.}
\end{figure}

Because of the closeness of the dust blob to Sgr\,A*, the two sources
could not be separated by {\it StarFinder}. In order to estimate the
flux of Sgr\,A*, we therefore proceeded in the following way. Using
the PSF extracted from the $M'$-images, two point sources were
subtracted from the images: one at the position of Sgr\,A* and one at
the position of the dust blob, $\sim0.094"$ SE of Sgr\,A*
\citep{Ghez:2005fk}. The flux of the sources was varied until, after
subtraction, the emission at the corresponding positions appeared flat
and equaled approximately the surrounding background (see
Fig.\,\ref{Fig:MpSgrA}). The best fit and the uncertainties were
estimated by eye.  We could not detect any significant extended
residuals after subtracting Sgr\,A* plus the dust blob from the
images. We therefore estimate that the assumption of a point-like
source for the dust blob is a good approximation.

In this way the flux density of the dust blob is estimated to be
$1.2\pm0.3$\,mJy. This agrees well with the $0.5-1.2$\,mJy that were
reported for its emission at $L'$.  The flux density of the point
source at the position of Sgr\,A* varies between $1.2$ and $1.8$\,mJy.
The individual measurements for Sgr\,A* are listed in
Tab.\,\ref{Tab:MpSgrA}. If no source at the position of the dust blob
were subtracted from the images, the measured flux densities of
Sgr\,A* would be about 0.3-0.4\,mJy higher. From this low
contamination and from some experimenting with variable positions and
flux densities for the dust blob, we estimate that any errors on the
assumed position and flux of the dust blob, including its
approximation as a point source, will result in systematic errors not
larger than about $0.1$\,mJy on the estimated flux density of Sgr\,A*.

Within the uncertainties there is no significant variability of the
putative $M'$-counterpart of Sgr\,A* between the observing
epochs. This does not exclude significant variability on timescales of
several 10 to 100 minutes as is reported for the $L'$-counterpart of
Sgr\,A* \citep[e.g.,][]{Ghez:2005fk,Eckart:2008tg,Dodds-Eden:2009mi}
and has also been found in the $M$-band
\citep{Hornstein:2007kx}. Here, we did not try to examine the
variability of Sgr\,A* at $M'$ on short timescales. Such observations
appear to be extremely difficult with NaCo. Chopping was used for all
observations. Due to the small chopping throw in the available data
and the crowded GC region we found that any reliable $M'$-image must
be composed of data at different dither positions. 
Chopping is not offered any more at the UT of the VLT where NaCo is
mounted.

A final question is whether the $M'$ emission from Sgr\,A* is
contaminated by emission from stars in the extremely dense cluster of
stars around the black hole. The stars in the so-called S-cluster are
mainly B-type main-sequence stars \citep{Gillessen:2009qe}.
\citet{Sabha:2010fk} report that unresolved stellar sources do not
contribute more than a reddened $Ks$-band flux density of 0.15\,mJy at
the position of Sgr\,A*. This corresponds to a mere $0.16$\,mJy in
$M'$ (see below for the extinction in $M'$). As concerns resolved sources, we can safely assume that no star
brighter than $Ks\approx15.5$ (about $0.3$\,mJy) was confused with
Sgr\,A* during the $M'$ observations \citep[inferred from stellar
orbits, see, e.g.,][]{Gillessen:2009qe}. We thus assign a conservative
$0.3$\,mJy as possible systematic $1\,\sigma$ uncertainty due to
contamination by stellar flux.

The time-weighted mean of all our measurements of the putative
  $M'$ counterpart of Sgr\,A* is $1.51$\,mJy.  The corresponding
  (time-weighted) uncertainty is $0.10$\,mJy. The systematic
  $1\,\sigma$ uncertainty due to possible contamination by stellar
  flux or from the dust blob is $0.3$\,mJy, as mentioned above. For
  the following analysis, we combine the statistical and systematic
  uncertainties in quadrature and obtain $0.32$\,mJy. Estimating
  $A_{M'}/A_{Ks}=0.4\pm0.05$ and its uncertainty from \citet{Nishiyama:2009oj}
  and using $A_{Ks}=2.46\pm0.03$ toward Sgr\,A* from \citet{Schodel:2010fk} we get
  $A_{M'}=1.0\pm0.3$. We thus obtain an extinction corrected flux
  density of $3.8\pm1.3$\,mJy for the $M'$ counterpart of Sgr\,A*,
  where the uncertainty of the extinction correction was taken into
  account.  The $M'$-flux density inferred in this work is similar to
  the one measured by \citet{Clenet:2004ve}, although they calibrate
  the zero point and extinction correction in a different way than in
  this work. \citet{Clenet:2004ve} do not discuss the
  influence of the nearby dust blob, which was only discovered in
  later work and is probably the cause of the astrometric offset of
  the Sgr\,A* source discussed by \citet{Clenet:2004ve}. The similar
  photometric results confirm the robustness of the photometry of
  Sgr\,A* in $M'$ and that the systematic uncertainty due to the
  nearby dust blob is not significant. Finally, as reported by
  \citet{Dodds-Eden:2011fk} for the $Ks-$band and shown below for the
  $L'-$band, Sgr\,A* spends only a small fraction of the time in high
  flux states. Therefore, sampling 8 random epochs \citep[9 when
  including the results of ][ see below]{Hornstein:2007kx}, will
  probably provide a reasonable estimate of the mean emission of
  Sgr\,A* in the $M'-$band. We note that the flux densities from
  the 8 VLT epochs agree within $\sim$$1\sigma$ of the individual
  uncertainties. 

\citet{Hornstein:2007kx} reported an extinction corrected flux density
of $7.3\pm1.7$\,mJy.  This difference can probably not be explained by
errors in calibration because our photometry agrees well with the one
of \citet{Hornstein:2007kx} (see section\,\ref{sec:MpObs}). Some of
the difference may be attributed to use of different extinction
laws. Applying the extinction law used by \citet{Hornstein:2007kx} to
our measurements, {\bf we obtain $5.8\pm2.0$\,mJy} for the extinction
corrected flux density from Sgr\,A*. Hence, the de-reddened flux
densities agree within their $1\,\sigma$ uncertainties. In this paper
we report the mean flux of Sgr\,A* averaged over a large data set
covering several epochs, while \citet{Hornstein:2007kx} report the
mean emission during just a single epoch. Sgr\,A* was probably more
active during their observations; in fact, the  light
  curve presented in \citet{Hornstein:2007kx} shows clear signs of ongoing
  activity. Since the total exposure time of their data is much longer
  than the one of the VLT data, including their value into our data
  set would mean that it completely dominates the statistics. We
  therefore refrain from mixing the two data sets because we do not know
  whether the event observed by \citet{Hornstein:2007kx} was typical
  or exceptional. Nevertheless, if we include their data, this changes
  the extinction-corrected mean $M'$ flux density of Sgr\,A* to
  $4.6\pm1.5$\,mJy, i.e. the difference is not significant. In
  any case, the independent observations by \citet{Hornstein:2007kx}
  support the notion that the mean flux of Sgr\,A* in the
  M'-band measured in the NaCo/VLT data is rather typical.

\subsection{The mean emission from Sgr\,A* at $3.8\,\mu$m \label{sec:L}}

Photometry on the $L'$-images from all available epochs was done with
{\it StarFinder}. The PSF was extracted by using 14 reference stars
within about $8"$ of Sgr\,A*. The correlation threshold was set to
0.7, and the $back\_box$ parameter was set to 2 pixels.

Two iterations were run on
each image with a $3\,\sigma$ noise threshold. Photometry was
calibrated by using the two stars $IRS\,16C$ and
$IRS\,16NW$\footnote{This star is classified as variable by
  \citet{Rafelski:2007fk}, but we consider the small reported variability negligible
for the purpose here. \citet{Ott:1999ly} did not report IRS\,16NW as variable.}
\citep[$L'=8.20, 8.43$][]{Schodel:2010fk}. 

\begin{figure}[!tb]
\includegraphics[width=\columnwidth,angle=0]{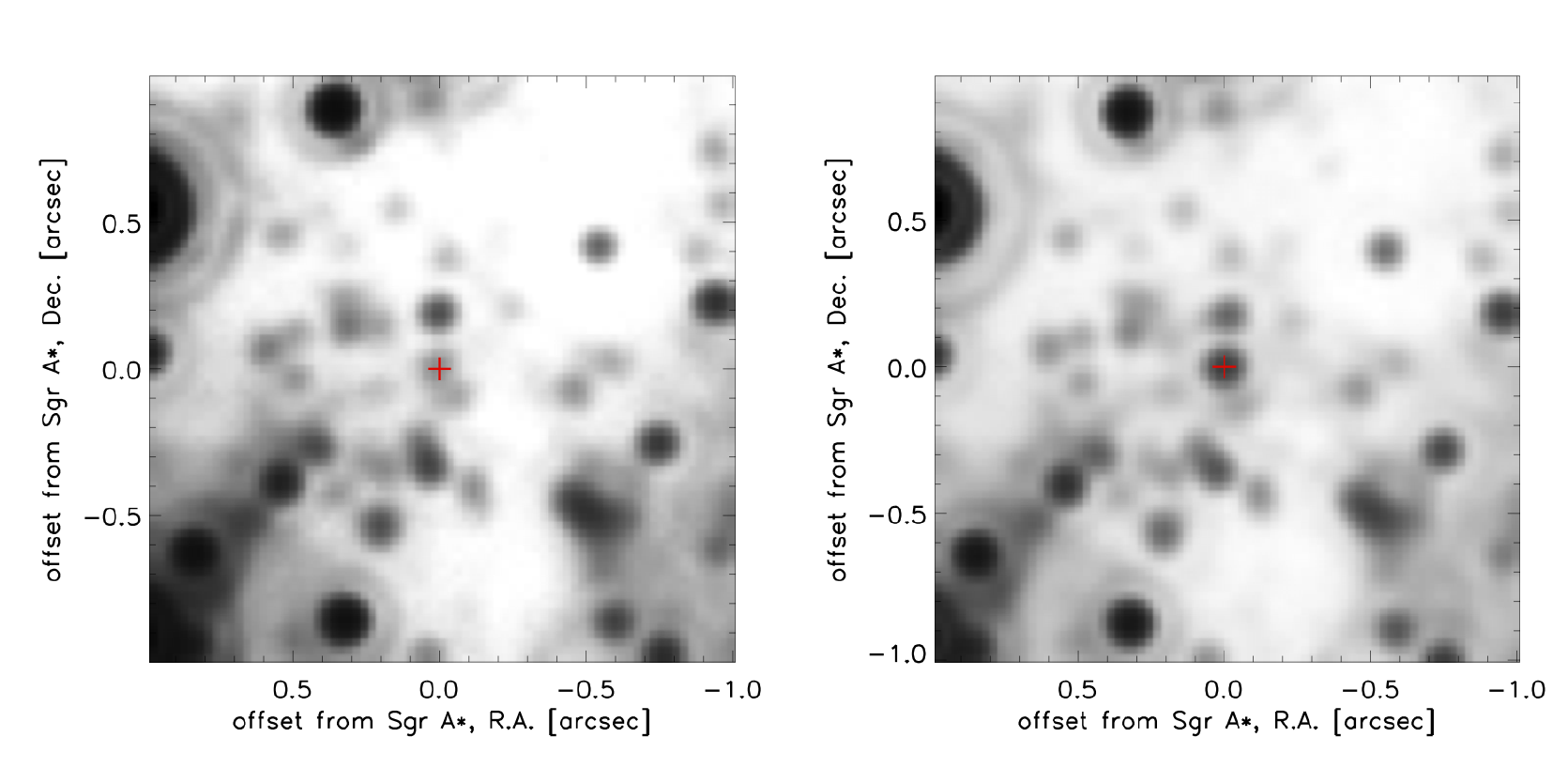}
\caption{\label{Fig:LpSgrA} Left: $L'$-image from 1 April 2007, one of
  the epochs with the faintest mean state of Sgr\,A*. Right:
  $L'$-image from 26 May 2008, one of the epochs with the brightest
  mean state of Sgr\,A*. The position of Sgr\,A* is indicated by a red cross.}
\end{figure}

The fourth and the sixth columns in Tab.\,\ref{Tab:LpObs} list the
measured fluxes of the star S2/S0-2 and of Sgr\,A*.  We believe that
contamination of the measured fluxes of Sgr\,A* and S2 by point
sources or by unresolved stars is negligible because of the high
Strehl ratio (estimated at $\gtrsim70\%$) of the $L'$-images and the
fitting of a finely sampled background emission. The unweighted
  means and the corresponding uncertainties of the flux densities are
  $f_{\rm S2} = 1.89\pm0.04$\,mJy (corresponding to $L' =
  12.79\pm0.02$) and $f_{\rm SgrA*} = 1.58\pm0.19$\,mJy (corresponding
  to $L' = 12.99\pm0.13$). The exposure-time-weighted means and
  corresponding uncertainties are $f_{\rm S2} = 1.90\pm0.02$\,mJy
  (corresponding to $L' = 12.79\pm0.02$) and $f_{\rm SgrA*} =
  1.62\pm0.20$\,mJy (corresponding to $L' = 12.96\pm0.13$).  No
variability of the star S2/S0-2 has been reported so far. Its measured
mean flux density is in excellent agreement with previously published
results \citep[e.g., $L'_{\rm S0-2} =
12.78\pm0.03$][]{Ghez:2005fk}. The dispersion of measured flux
densities of SgrA* is greater by a factor of almost 20 than that of
S2/S0-2, which clearly indicates the variability of this source. In
the following, we will use the uncertainty of the mean emission of
Sgr\,A*, which is $0.20\,$mJy, and not the standard deviations
calculated from the individual measurements. $L'$-images of the close
surroundings of Sgr\,A* are shown for two epochs in
Fig.\,\ref{Fig:LpSgrA}.

After applying an extinction correction of $A_{\rm L'}=1.23\pm0.08$
\citep{Schodel:2010fk}, we obtain a de-reddened mean flux density of
$5.0\pm0.6$\,mJy for Sgr\,A* (uncertainty of the extinction
correction was included in the uncertainty of the flux density).

In order to test whether the mean flux density is sensitive to
  rare events, like exceptionally bright and long flares
  \citep[e.g.,][]{Dodds-Eden:2009mi,Kunneriath:2010fk}, the flux
  density of Sgr\,A* was also measured on the median images of each
  epoch. As can be seen in column 5 of Tab.\,\ref{Tab:LpObs}, the
  values are generally slightly lower, but hardly different from the
  values measured on the mean images. 

  Finally, we created subsets of the data from each epoch in order to
  construct 74 independent images from the existing data, of about
  800\,s exposure time each.  The 800\,s exposure time offers a good
  compromise between a relatively high time resolution, on the one
  hand, and the need to average a sufficient number of frames in order
  to create clean images, on the other hand. This is particularly
  important in the case of L'-imaging because of the high thermal
  background and the difficulty of extracting accurately the
  instantaneous sky emission from a small number of dithered frames on
  an extended and bright target. 

\begin{figure}[!tb]
\includegraphics[width=\columnwidth,angle=0]{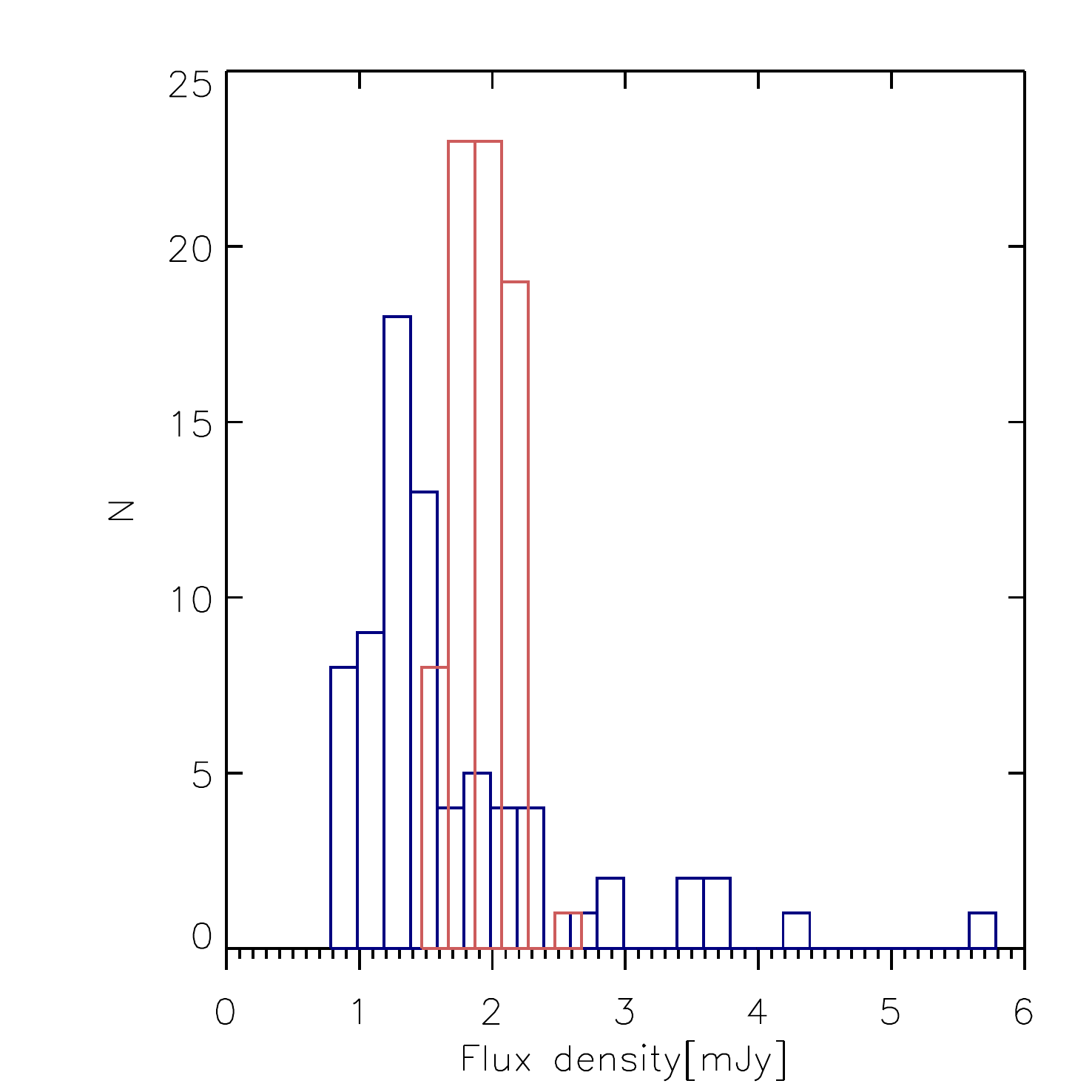}
\caption{\label{Fig:800s} Blue: Histogram of measured L'-flux densities
  (not corrected for extinction) of Sgr\,A* in 800\,s exposures. Red:
  Equivalent histogram for the constant star S2.}
\end{figure}

  The 800\,s-images were inspected by eye and analyzed with {\it
    StarFinder}. Sgr\,A* is detected in each image. The measured flux
  distributions of Sgr\,A* and of S2 are shown in
  Fig.\,\ref{Fig:800s}.  The histogram of the flux densities of the
  constant comparison star S2 is close to Gaussian and provides an
  estimate of the photometric uncertainty. The histogram of Sgr\,A* is
  significantly broader, with a tail toward high flux densities,
  similar to what has been found in a detailed analysis of the
  variability of Sgr\,A* in the $Ks-$band \citep{Dodds-Eden:2011fk}.
  It is beyond the aims of this paper to discuss the exact
  time-variable behavior of Sgr\,A* in the $L'$-band. Here, it is only
  important that Sgr\,A* is always detectable on images with
  integration times on the order of 13 min. It appears to spend most
  of the time in a low state between about $1-2$\,mJy (not corrected
  for extinction), and shows only occasional excursions toward high
  flux values, potentially reflecting overlapping {\it quiescent} and  {\it
    flaring} states \citep[see][]{Dodds-Eden:2011fk}.  For example,
  the $3.8\,\mu$m flux density of Sgr\,A* exceeds $2.0\,$ ($2.5$)\,mJy
  (not corrected for extinction) in only 23\% (12\%) of the
  800\,s-images. This underlines that it is meaningful to speak of a
  mean flux density of Sgr\,A* in the L'-band. 

It appears that, in the $L-$band, Sgr\,A* is continually variable on
levels between 1 and several 10 mJy \citep[all flux densities
extinction corrected,
e.g.,][]{Ghez:2005fk,Eckart:2008tg,Dodds-Eden:2009mi}. Several long
observations with the Keck telescope and corresponding mean flux
densities of Sgr\,A* are reported in \citet{Do:2009ij} and
\citet{Hornstein:2007kx}. From their results, after correction for the
$L'$-extinction used in this work, we obtain an exposure-time-weighted
mean de-reddened $L'$-flux density of Sgr\,A* of
  $11.8\pm2.6$\,mJy. This is about a factor of 2 higher than the mean
emission estimated from the NaCo $L'$ imaging data. However, also the
uncertainty of the mean is significantly higher, so that the mean flux
density agrees with the NaCo flux density within about
$2\,\sigma$. The mean flux density from the Keck
observations is based on only 5 epochs or about 8\,h of observing
time, while the mean flux from the NaCo results is based on 19 epochs
or roughly 19\,h. Therefore the $L'$ mean flux density from the Keck
results may be biased by a few exceptionally bright events, which is
supported by the high standard deviation of the flux densities
reported for the Keck observations.

\subsection{The mean emission from Sgr\,A* at $2.1\,\mu$m \label{sec:K}}

The stellar density in the GC is highest near Sgr\,A* and confusion of
the very red counterpart of Sgr\,A* with stellar sources becomes an
increasingly serious problem in the $K$ and $H$ bands \citep[see,
e.g.,][]{Hornstein:2007kx}, particularly when Sgr\,A* is in a faint
state.

Since it is difficult to disentangle Sgr\,A* in a low flux state from
the faint nearby stars \citep[for a good example, see Fig.\,1
in][]{Sabha:2010fk}, work on the variable emission from Sgr\,A* based
on VLT observations was, for a long time, focused almost exclusively
on flaring emission. Only recently, \citet{Dodds-Eden:2011fk}
presented an extensive and detailed study of the emission of Sgr\,A*
in the $Ks-$band with NaCo/VLT. Based on a large number of
observational epochs, they present a lightcurve with an equivalent
length of $\sim184\,h$. Although they cannot separate Sgr\,A* from a
stellar source, S17, that is almost coincident with Sgr\,A* during
most epochs, they provide a detailed estimate of the amount of
contamination of the emission from Sgr\,A* by stellar light. Their
main conclusions relevant to this work are that Sgr\,A* is a
continuously variable source in the $Ks-$band and that it spends the
overwhelming majority of its time in a low variability state. This low
state is termed the {\it quiescent state} by the authors and is
characterized by a log-normal flux distribution. They derive a
  de-reddened median flux of $1.6$\,mJy at the position of of Sgr\,A*, $0.5$\,mJy
  of which is attributed to stellar contamination and $1.1\,$mJy to
  Sgr\,A* itself.

  At times, Sgr\,A* can apparently not be detected with NaCo/VLT. For
  example, \citet{Sabha:2010fk} analyze deep, high-quality imaging
  data from NaCo during a low phase of Sgr\,A*. In spite of their
  imaging data being among the best $Ks-$band GC imaging data obtained
  at NaCo/VLT (in terms of Strehl ratio and stability of AO
  correction), they do not detect any Sgr\,A* counterpart and give a
  de-reddened upper limit of $1-1.5$\,mJy (extinction corrected for
  the extinction values assumed in this
  work). \citet{Dodds-Eden:2011fk} estimate, however, that such
  extremely low states are relatively rare ($\sim16\%$ of the time). 

The Keck telescope, on the other hand, is less prone to confusion
because of its larger aperture and consequently higher angular
resolution. Therefore, work on the emission from Sgr\,A* based on Keck
data has not been biased toward flares or brighter states of
Sgr\,A*. The relevant publications report that Sgr\,A* is detected in
the $K'$-band ($2.12\,\mu$)-band at all times
\citep{Hornstein:2007kx,Do:2009ij,Meyer:2009tg}. Table\,1 in
\citet{Meyer:2009tg} lists mean $K$-band\footnote{The Keck
    measurements are in $K'$, with a central wavelength of
    $2.1\,\mu$m, while the VLT measurements are at $Ks$ with
    $\lambda_{c}=2.18\,\mu$m; here, we neglect this difference and
    assume roughly $\lambda_{c}=2.1\,\mu$m also for the two VLT
    measurements.} flux densities of Sgr\,A* measured with the Keck
telescope during 12 epochs, with one mean flux density from combined
Keck/VLT data sets. They also list a mean flux density from one
VLT-only observing run. An inspection of the relevant lightcurves in
\citet{Eckart:2006sp} and \citet{Meyer:2008fk} shows that Sgr\,A* was
detected during the cited VLT observations all the time. Here, we
exclude the $L'$ and the mixed $L'/K$-band data sets \citep[first two
lines in Table\,1 of][]{Meyer:2009tg}. The measurements listed in
\citet{Meyer:2009tg} are de-reddened with an extinction of
$A_{K}=3.2$\,mag. We removed this reddening correction and calculated
the mean flux density of Sgr\,A* in the $K$-band, weighted by the
given length of the observations.  It is
$0.20\pm0.02$\,mJy\footnote{The unweighted average and standard
  deviation are $0.208\pm0.055$\,mJy}, corresponding to a source of
$K=16.24\pm0.11$.  The quality of the NIRC/Keck $K$-data is high and
point-source fitting works well. Also, {\it StarFinder} estimates a
diffuse background in parallel with point source fitting. This should
largely suppress any contamination from unresolved stars, so that
  a large fraction of the flux measured at the position of Sgr\,A*
  will in fact be related to this source, with stellar contamination
  estimated to be $\lesssim35\%$ by \citet{Do:2009ij}.  Correction for
  $A_{K}=2.59\pm0.03$\,mag of extinction \citep[assuming a
    central wavelength of $2.1\,\mu$m and following the extinction
  law given in][]{Schodel:2010fk} results in a de-reddened flux
  density of $2.2\pm0.2$\,mJy at $2.12\,\mu$m (uncertainty of the
  extinction correction included). 

 As can be seen, the situation in the $K-$band is more complicated
  than in the $L'-$ and $M'-$bands. Particularly, disentangling
  Sgr\,A* from the surrounding stars almost coincident in position
  with the black hole, requires a very careful analysis and the
  highest possible angular resolution and image quality. Observing a
  clearly isolated point source at all times, even in its faintest
  states, will probably require larger telescope apertures, although
  the aperture of the Keck telescope appears to be marginally sufficient
  for this purpose. In any case, the works of \citet{Do:2009ij} and
  \citet{Dodds-Eden:2011fk} show convincingly that Sgr\,A* is
  permanently detectable in the $K-$band with a flux in the range
  $0.5-2.5$\,mJy.

\section{The IR emission from Sgr\,A* in the context of its SED\label{sec:models}}

\begin{figure}[!htb]
\includegraphics[width=0.8\columnwidth,angle=0]{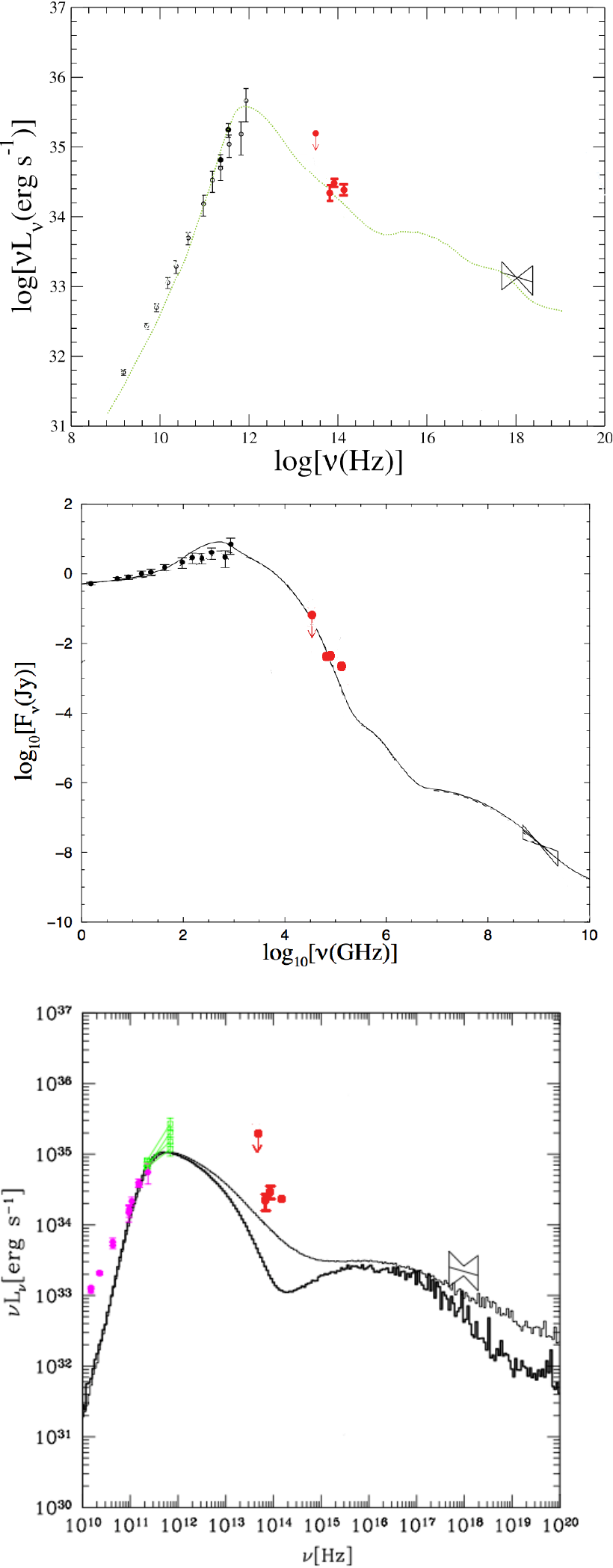}
\caption{\label{Fig:models} SED models for the quiescent or mean
  emission from Sgr\,A*. Top: RIAF model by
  \citet{Yuan:2004fk}. Middle: Jet-ADAF model by
  \citet{Yuan:2002uq}. Bottom: SED from a general relativistic
  magnetohydrodynamics simulation by \citet{Moscibrodzka:2009kx}
  (their ``best-bet'' model is shown), where the thin line is the
  average SED and the thick line represents the SED at a given moment
  during the simulations. The upper limit at $8.59\,\mu$m
  is indicated by a down-pointing red arrow, the mean flux densities at
  $4.8$, $3.8$ and $2.1\,\mu$m are shown as red dots with error bars,
  which correspond to the standard deviation of the observed
  variability between epochs.  For details on the models and
  radio/mm/X-ray data, see the referenced publications.}
\end{figure}

 Work on the infrared emission of Sgr\,A* has so far been focused
  mainly on the flaring emission, not on the apparently less exciting
  mean emission.  As we have seen, in the sections above, imaging data
  provide evidence that Sgr\,A* is detected at $4.8$, $3.8\,\mu$m, and
  $2.1-2.2\,\mu$m {\it at all times}.  Moreover, the mean flux of
  Sgr\,A* is well defined at these infrared wavelengths. Although the
  timescales analyzed at the different wavelengths are somewhat
  different, from a few to to 13 min in case of $M'$ and $L'$ and
  subminute timescales in case of $K$, these differences should not
  be important for the mean SED of Sgr\,A*, which emphasizes the
  activity of Sgr\,A* over long timescales (days to years). Since this
  paper is focused on the mean NIR/MIR emission from Sgr\,A*, we are
  not concerned with the rapid variability of the emission of Sgr\,A*
  on minute time scales. 

Strictly speaking, the mean of a stochastic process (and the flux from
Sgr\,A* can be described in terms of a stochastic process) is only
time independent for a (weakly-) stationary process. As the power
spectrum of SgrA* follows a broken power-law with a slope for low
frequencies consistent with zero and a break timescale of ~160 min
\citep{Meyer:2009tg}, it can be expected that the flux's mean is
time-independent for averaging times greater than ~160 min. This holds
true for most of our observations, in particular for the mean flux
densities, considered here, which result from an average over a large
number of epochs. The analysis in this section implies that there is
no significant trend of the mean emission from Sgr\,A* over the time
range covered in this work. The $L'$ and $M'$ data in this work do not
provide any evidence for such a trend. Some evidence for
  variations on time scales of weeks to months is found in the
  $K-$band flux density measurements by
  \citet{Dodds-Eden:2011fk}. However, so far, there is no evidence for
  any overall increasing or decreasing long-term trend.

Since it has not been clear for many years whether the infrared
  counterpart of Sgr\,A* can only be detected during flares and
  assumes a kind of {\it off}-state in between flares, there may still
  exist some lack of clarity in the community about the term {\it
    quiescence}.  \citet{Eckart:2006sp} referred to a state of
  "low-level, and, especially in the NIR domain, possibly continuously
  variable flux density'' as the {\it interim-quiescent} state.
  \citet{Sabha:2010fk} use the term to refer to a state in which
  Sgr\,A* cannot be detected because the images and analysis presented
  in their work show clearly how difficult, even impossible, it is to
  detect  Sgr\,A* on NaCo/VLT images when it is in a very faint
  state. \citet{Dodds-Eden:2011fk}, however, measured the emission of
  Sgr\,A* plus underlying stellar source(s) at the position of Sgr\,A*
  in an extensive data set. They estimate in detail the stellar
  contribution to the flux of Sgr\,A* in the $Ks-$band. They define
  two states by separating the distribution of measured $Ks-$fluxes
  into a log-normal and a power-law component and refer to the
  log-normal state of low-variability as the {\it quiescent state}.

  The distribution of $L'-$band fluxes found in this work (see
  Fig.\,\ref{Fig:800s}) shows that Sgr\,A* is always clearly
  detectable at $3.8\,\mu$m, and that it spends much time in a
  low-variability state, with a tail toward high flux densities,
  similar to what has been found in the more detailed analysis of
  \citet{Dodds-Eden:2011fk} for the $Ks$-band.  The mean flux density
  derived for Sgr\,A* at $L'$ is therefore well-defined and probably
  reflects the quiescent state as defined by
  \citet{Dodds-Eden:2011fk}.  Confusion with stellar sources and
  contamination by stellar flux, respectivly, are of minor importance
  at $3.8\,\mu$m as compared to $2.2\,\mu$m \citep[see, e.g.,
  discussion in][]{Do:2009ij}. This underlines the importance of the
  $L'$ data in the context of the mean SED of Sgr\,A*.

  In the MIR, at a wavelength of $8.6\,\mu$m, we find a
    de-reddened $3\,\sigma$ upper limit of $84$\,mJy on the quiescent
    flux density of Sgr\,A*. This is consistent with, but lower than
    in previous findings \citep{Schodel:2007hh} at this wavelength,
    when differences in extinction estimates and their uncertainties
    are taken into account.  We believe that the improvement is based
  on the use of imaging data of unprecedented quantity and
  quality. We take the uncertainty of the extinction
  correction explicitly into account when estimating the $3\,\sigma$
  upper limit. \citet{Dodds-Eden:2009mi} report a comparable
    de-reddened $3\,\sigma$ upper limit of 86\,mJy at
    $11.88\,\mu$m\footnote{This is higher than the $57$\,mJy reported
      in their work, because they do not explicitly take into account
      the uncertainty of the extinction correction for their upper
      limit. The $86$\,mJy are their $3\,\sigma$ upper limit when
      including the latter source of uncertainty.}.

We summarize our measurements of the upper limit of the emission at
$8.6\,\mu$m and the mean emission at $3.8\,\mu$m and $4.8\,\mu$m of
Sgr\,A* in Tab.\,\ref{Tab:SgrAIR}, where we also list the mean
$K$-band flux densities based on the work of \citet{Meyer:2009tg} and
\citet{Dodds-Eden:2011fk}. We list the (exposure-time-weighted)
uncertainties of the mean flux densities, and not the
(exposure-time-weighted) standard deviations because we are focused on
the mean SED from Sgr\,A*. The standard deviations are useful for giving
an impression of the epoch-to-epoch variability of Sgr\,A*. At 
$3.8\,\mu$m ($4.8\,\mu$m) they are a factor of $\sim$$4$
($\sim$$2.5$) higher than the uncertainties of the mean flux
densities.

\begin{table}[htb]
  \caption{De-reddened mean flux densities of Sgr\,A* in the infrared. \label{Tab:SgrAIR}} 
%\centering
\begin{tabular}{lll}
\hline
\hline
$\lambda$ & f & Extinction\tablefoottext{a}  \\
($\mu$m)  & (mJy) & (mag) \\
\hline
8.6 & $84$\tablefoottext{b} & $2.0\pm0.3$\\
4.8 & $3.8\pm1.3$ & $1.0\pm0.3$\\ 
3.8 & $5.0\pm0.6$ & $1.23\pm0.08$\\
2.1\tablefoottext{c} & $2.2\pm0.2$ & $2.59\pm0.03$\\
2.18\tablefoottext{d} & $1.1$ & $2.5$\\
\hline
\end{tabular}
\tablefoot{
  \tablefoottext{a}{Assumed extinction at this wavelength.}
  \tablefoottext{b}{De-reddened $3\,\sigma$ upper limit, including
    the uncertainty of the extinction correction. The {\it
        measured} $3\,\sigma$ upper limit, i.e., not corrected
      for extinction, is $10$\,mJy.}
  \tablefoottext{c}{Calculated from the data given in \citet{Meyer:2009tg}.}
  \tablefoottext{d}{Median flux density from \citet{Dodds-Eden:2011fk}. They give a
      multiplicative standard deviation of $2.1$ for the median flux.}
}
\end{table}

Up to now, infrared observations have almost exclusively played a role
for modeling the flaring emission from Sgr\,A*. For models of the mean
emission, the infrared measurements have so far been used only in the
form of upper limits. Various models for the SED of Sgr\,A* are shown
in Fig.\,\ref{Fig:models}, along with measurements at radio and X-ray
wavelengths as well as the new infrared measurements and upper
limit from this work. It is clear that the infrared measurements can
provide reliable anchor points for the mean SED of Sgr\,A* on
the high-frequency side of the submillimeter bump, providing
measurements in a previously existing gap on the order of 6 magnitudes. 

The RIAF and the jet-ADAF models shown in Fig.\,\ref{Fig:models}
appear to fit the mean infrared fluxes/flux densities and upper limits
in general well. The MIR upper-limit lies close to the predicted
  flux in the jet-ADAF model.  The new data
points can probably help to fine-tune some parameters of these models.

The GRMHD model from \citet{Moscibrodzka:2009kx} appears to
under-predict the IR flux densities. However, we show just
one of their many models, i.e, what they call the ``best-bet''
model. From the models shown in Fig.\,4 of their work, it appears that
the new infrared measurements generally favor those models (with
  $T_{i}/T_{e}=1$ or $3$) having high angular momentum of the black hole
  ($a_{*}\gtrsim0.95$) and high inclination angle ($\sim$$85\deg$) of
  the accretion flow. However, those models appear to become only
  marginally consistent with the X-ray constraints. 

The measured mean emission at $2.1$ and $3.8\,\mu$m and the upper
limit at $8.6\,\mu$m show a clearly increasing trend, in agreement
with the models. The $4.8\,\mu$m point, however, appears to be too
low, and not follow the trend. The discrepancy is not major,
considering the uncertainties of the measurements. It may be possible
that the $M-$band data are biased by insufficient sampling. One has
to recall that, due to the low efficiency, the entire NaCo $M'$ data
set only comprises a total on-source integration time of just about
650\,s. Nevertheless, the mean IR flux densities may also indicate a
somewhat flatter slope of the SED in this regime than what is
predicted by the models. Therefore this point may merit additional
measurements.

\section{On the detectability of Sgr\,A* in the MIR}

During flares the emission from Sgr\,A* has been reported to rise as
high as a few tens of mJy in the NIR
\citep[e.g.,][]{Dodds-Eden:2009mi,Kunneriath:2010fk}. However, Sgr\,A*
has so far never been detected in the MIR, not even during flaring
activity \citep{Schodel:2007hh,Dodds-Eden:2009mi}. What can we
reasonably expect with VISIR/VLT or similar telescope-instrument
combinations?

A considerable amount of observing time has been invested with
VISIR/VLT to detect Sgr\,A*. The sensitivity of VISIR with the PAH1
filter allows the detection of a 5\,mJy point source with $10\,\sigma$
significance in 1 hour of on-source integration time. These
sensitivity estimates were derived from tests on standard stars with
chopping and nodding done in a way that the target was always within
the detector FOV (see ESO VISIR user's manual). The GC is an extended
target, so chopping and nodding have to be done off-source, lowering
the observing efficiency to 25\% and thus the 1\,hour sensitivity
limit by a factor $2$. Due to the limited chop-angle and
imperfections on the VISIR detector, some dithering and/or change in
chop-angle is also highly recommended (and is usually applied). This
will further decrease the observing efficiency. Burst mode
observations have an even lower efficiency, about 7\% (estimated from
our data), but we judge that, for point sources, the perfect image
quality delivered by the holography technique largely offsets the loss
in efficiency. Thus, we assume similar sensitivity limits for the
burst mode as for the standard observing technique. The theoretical
$\sim10$\,mJy (at 10\,$\sigma$) limit that can thus be reached under
ideal observing conditions in 1\,hour of observing time (corresponding
to 15 min on source) is very close to the $3\,\sigma$ upper limits
derived from our observations. The uncertainty of the upper limit on
Sgr\,A* reported in this work is higher than what could be expected
from the theoretical sensitivity, particularly considering the long
exposure times of some of the presented images. However, this is
because Sgr\,A* is confused with a ridge of MIR emission and our lack
of any image where we know with certainty that the emission from
Sgr\,A* can be neglected, which force us to make conservative
assumptions and result in these higher uncertainties.

If we consult the theoretical SED models of Sgr\,A* that were
presented in the previous section, we see that the RIAF model predicts
an $8.6\,\mu$m luminosity about 6 times lower than the current
$3\,\sigma$ upper limit. The jet model, on the other hand, suggests
that Sgr\,A* may linger near the detection limit at MIR
wavelengths. Deeper MIR observations may therefore possibly serve to
discriminate between pure RIAF and RIAF+jet models. Of course, the
models will first have to be updated with their parameters adjusted to
the new infrared constraints. Nevertheless, we estimate that the
sensitivity of the MIR instrumentation should be increased about
five- to ten-fold in order to guarantee significant progress.

Can Sgr\,A* be detected in the MIR during a flare? The brightest
states of the infrared flares, during which detection in the MIR can
be expected to be most likely, last normally about 20\,min. These are the
so-called subflares \citep[e.g.,][]{Eckart:2008fk}. For detecting
flares, the time resolution of the MIR observations should be of the
same order. Due to the large overheads of MIR observations of the GC,
this means, however, that with such a time resolution it is hardly
possible to apply dithering and/or to use different chopping positions
in order to minimize systematic errors caused by detector defects or
the extended flux of the GC region. The data from 22/23 May 2007
should be largely free of such effects, however.\footnote{The imaging data from
  May 2007 were observed by R. Schoedel, not downloaded from the
  archive, i.e. we can be sure of the set-up. Great care was taken to
  avoid chopping into extended sources.} From these data we produced
five images reconstructed from 4000 different frames each, corresponding to
about 19 min time resolution and 80\,s on-source integration for each
reconstructed image.

\begin{figure}[!tb]
\includegraphics[width=\columnwidth,angle=0]{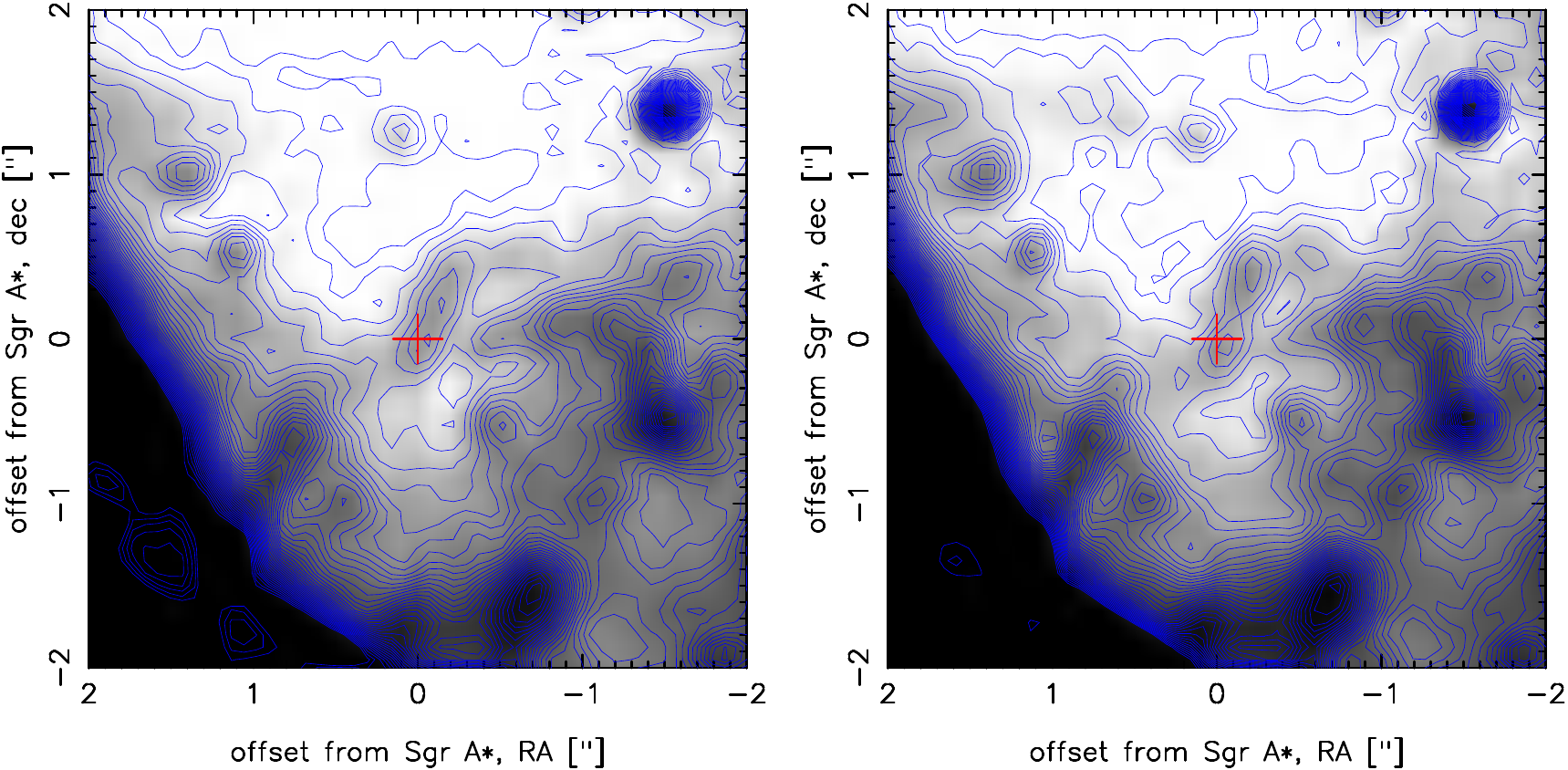}
\caption{\label{Fig:variability} Two images at $8.6\,\mu$m,
  reconstructed from subsets of the May 2007 data. Each image is based
  on a different set of 4000 frames, corresponding to an on-source
  integration time of 80\,s during an interval of about 19\,min. The
  location of Sgr\,A* is marked by a red cross. The apparent
  variability of some of the extended features is caused by the low
  S/N of these images. The contours are as in
  Fig.\,\ref{Fig:overview}.}
\end{figure}

On these images the flux density in a circular aperture of $0.225"$
{\bf radius} was measured in an empty region about $0.5"$ NE of
Sgr\,A*. The $1\,\sigma$ uncertainty from the measurements was
$\sim$7\,mJy. Additionally, the flux density within an identical
aperture was measured on the same images at the position of Sgr\,A*
after subtracting from them the average image reconstructed from all
May 2007 data. The inferred $1\,\sigma$ uncertainty at the position of
Sgr\,A* was $\sim5$\,mJy, i.e., close to the one of the empty field
(and clearly not larger, as would be expected if Sgr\,A* had
varied). Two of the 20 min snapshots from the May 2007 data are shown
in Fig.\,\ref{Fig:variability} in order to give the reader an
impression of the random variability of the faint features near
Sgr\,A* on images with such short integration times.

Hence, to detect Sgr\,A* with $5\,\sigma$ significance during a
20\,min time interval, the source must be on average as bright as
$\sim$30\,mJy ($\sim$190\,mJy extinction corrected). This is, for
example, a factor of $>10$ above the quiescent emission predicted by the
model shown in the upper panel of Fig.\,\ref{Fig:models}. 

From NIR measurements it has been inferred that Sgr\,A* has an
approximately constant spectral index of $\alpha=-0.6\pm0.2$
\citep[for $f\propto\nu^{\alpha}$,][see also M. Bremer, in
prep.]{Gillessen:2006fk,Hornstein:2007kx} for flares with de-reddened
flux densities $\gtrsim5$\,mJy in the $K$-band. Assuming that this
spectral index can be extrapolated to the MIR, a flare must be
brighter than about $80$\,mJy (de-reddened) in the $K-$band
during at least 20\,min in order to be detected with VLT/VISIR at
$8.6\,\mu$m. To the best of our knowledge, such a bright flare has
never been reported. The situation would be more favorable if the
NIR-to-MIR spectral index is steeper, e.g.  $\alpha_{\rm
  K-L}=-1.4\pm0.3$ from the mean flux densities at $K'$ (Keck data)
and $L$ derived here. In that case, it would be sufficient if the
  flare were brighter than about $24$\,mJy (de-reddened) in the
  $K-$band during at least 20\,min.  Such a steep spectral index is
  probably an extreme assumption, and such bright flares appear to be
  extremely rare, which can, for example, be seen clearly in
  compilations of long time series of the emission from Sgr\,A*
  \citep[e.g.,][]{Meyer:2008fk,Meyer:2009tg,Yusef-Zadeh:2009fk,Dodds-Eden:2011fk}.
  In fact, this criterion is only met by a single flare reported in
  the literature \citep{Dodds-Eden:2011fk}.

It may still be possible that Sgr\,A* can be detected in the MIR
during exceptional events, such as the brightest $L-$band flare
  ever reported, during which Sgr\,A* showed flux densities between 60
  to 90\,mJy at $L'$ during a full hour
  \citep{Kunneriath:2010fk}. However, such events appear to be
  exceptionally rare. Here it is important to note that Sgr\,A* was
  not even detected in VISIR/VLT observations during the extremely
  bright, soft-spectrum X-ray flare reported by \citet{Porquet:2008kx}
  \citep[see ][for the simultaneous MIR
  observations]{Dodds-Eden:2009mi}

The analysis presented here is for $8.6\,\mu$m.  Similar conclusions
can be drawn for longer wavelengths, where the increasing thermal
background and decreasing angular resolution present additional
complications.  To summarize, we believe that Sgr\,A* cannot be
detected with VLT/VISIR or similar telescope-instrument combinations,
with the possible exception of rare extreme events. In order to detect
Sgr\,A* and examine its variability in the MIR, one needs to combine
the highest possible angular resolution with significantly
improved detectors. Possibly, very sensitive polarimetric observations
could reveal a Sgr\,A*-counterpart during flares \citep[as for the
polarization of Sgr\,A* in the NIR, see, e.g.,][]{Eckart:2006vn}.  An
alternative may be measurements under the extremely stable and
low-background conditions in space. The angular resolution of the JWST
will be required for this purpose.

\section{Summary}

This paper presents new observational data on the mean infrared
emission from Sgr\,A*. No counterpart could be detected at
$8.6\,\mu$m, in spite of using an extensive data set and images with
excellent Strehl ratios. Detection of a point-source at the location
of Sgr\,A* is complicated considerably by the presence of a ridge-like
structure, the {\it Sgr\,A*-Ridge}.  The $3\sigma$ upper limit on the
flux density from Sgr\,A* at $8.6\,\mu$m was estimated to be 10\,mJy
(observed) and 84\,mJy after de-reddening. This is lower than what has
been previously reported at this wavelength. The upper limit at
$8.6\,\mu$m is mainly dominated by the relatively high and uncertain
extinction at this wavelength.  Based on the sensitivity of existing
imaging data, we argue that MIR emission from Sgr\,A* -- both
time-averaged as well as flaring emission -- can probably not be
detected by imaging observations with current telescopes and
instruments. We estimate that about a ten-fold increase in
point-source sensitivity is needed in order to provide significant new
constraints.

At the shorter wavelengths of $4.8\,\mu$m and $3.8\,\mu$m, we
  find from an analysis of an extensive data set that a counterpart of
  Sgr\,A* can be detected at all times, and derive its mean
  emission. Finally, we use/derive the mean emission of Sgr\,A* at
  $2.1-2.2\,\mu$m from published Keck/VLT data sets, where it has
  recently been shown to be detectable at all times, too. 

The new infrared measurements are in general agreement with current
models for the SED of Sgr\,A*.  They do not allow us to clearly
distinguish between various published models, but will clearly help
to fine-tune the parameters of those models. It should be pointed out
that, so far, infrared data have not been included in the models, or,
at most, in the form of weak upper limits in the mid-infrared. We can
now be sure that Sgr\,A* is always detectable on the high-frequency
side of the Terahertz peak, with a rather well constrained mean
flux. The $8.6\,\mu$m upper limit may be low enough to have an
  appreciable impact on some models.  The data point at $3.8\,\mu$m is
  particularly well defined because it is based on a large data set,
  suffers negligible contamination by stellar light, and the
  uncertainty of the extinction at this wavelength is low.

\begin{acknowledgements}
  RS acknowledges support by the Ram\'on y Cajal programme, by grants
  AYA2010-17631 and AYA2009-13036 of the Spanish Ministry of
  Science and Innovation, and by grant P08-TIC-4075 of the Junta de
  Andaluc\'ia. AA acknowledges support by grant AYA2009-13036 of the
  Spanish Ministry of Science and Innovation and by grant P08-TIC-4075
  of the Junta de Andaluc\'ia.
\end{acknowledgements}

\bibliography{/Users/rainer/Documents/BibDesk/BibGC}

\end{document}